\documentclass[11pt]{article}
\usepackage{amsmath,amssymb,color,epsf}
\usepackage{amssymb,slashed,latexsym,amsmath,multirow,color,dsfont}
\usepackage{graphics}

\makeatletter
\@addtoreset{equation}{section}
\makeatother

\usepackage{amsfonts}

\usepackage{float}

\usepackage{soul}

\usepackage{amsfonts}

\usepackage[font=footnotesize,labelsep=newline,labelfont=sc,justification=centering,position=top]{caption}

\newcommand{\bea}{\setlength\arraycolsep{2pt} \begin{eqnarray}}
\newcommand{\eea}{\end{eqnarray}}
\newcommand{\nn}{\nonumber}

\usepackage{hyperref}

\newcommand{\ft}[2]{{\textstyle\frac{#1}{#2}}}
\def\rmi{{\rm i}}
\newsavebox{\uuunit}
\sbox{\uuunit}
    {\setlength{\unitlength}{0.825em}
     \begin{picture}(0.6,0.7)
        \thinlines
        \put(0,0){\line(1,0){0.5}}
        \put(0.15,0){\line(0,1){0.7}}
        \put(0.35,0){\line(0,1){0.8}}
       \multiput(0.3,0.8)(-0.04,-0.02){12}{\rule{0.5pt}{0.5pt}}
     \end {picture}}

\def\be{\begin{equation}}
\def\ee{\end{equation}}
\def\ba{\begin{array}}
\def\ea{\end{array}}
\def\bea{\begin{eqnarray}}
\def\eea{\end{eqnarray}}
\def\bd{\begin{displaymath}}
\def\ed{\end{displaymath}}

\def\nn{\nonumber}

\def\g{\gamma}

\def\d{\delta}

\def\e{\epsilon}
\def\ve{\varepsilon}

\def\vf{\varphi}

\def\p{\psi}

\def\l{\lambda}

\def\m{\mu}
\def\n{\nu}
\def\r{\rho}
\def\s{\sigma}

\def\o{\omega}
\def\O{\Omega}

\def\nn{\nonumber}
\def\cD{\mathcal{D}}
\def\cN{\mathcal{N}}

\def\cL{\mathcal{L}}
\def\cF{\mathcal{F}}

\newcommand{\eq}[1]{(\ref{#1})}
\newcommand{\w}[1]{\\[0.#1cm]}

\begin{document}

\begin{titlepage}

\begin{center}

\hfill UG-14-19 \\  \hfill MIFPA-14-37

\vskip 1.5cm

{\Large \bf Massive $\cN=2$ Supergravity in Three Dimensions}
\vskip 1cm

{\bf G{\"{o}}khan Alka\c{c}\,$^1$, Luca Basanisi\,$^1$, Eric A.~Bergshoeff\,$^1$, \\[0.5ex] Mehmet Ozkan\,$^1$ and Ergin Sezgin\,$^2$ } \\

\vskip 25pt

{\em $^1$ \hskip -.1truecm Van Swinderen Institute for Particle Physics and Gravity,  \\ 
University of Groningen, Nijenborgh 4, 9747 AG Groningen, The Netherlands \vskip 5pt }

{email: {\tt g.alkac@rug.nl, l.basanisi@rug.nl, e.a.bergshoeff@rug.nl, m.ozkan@rug.nl}} \\

\vskip 15pt

{\em $^2$ \hskip -.1truecm George and Cynthia Woods Mitchell Institute for Fundamental Physics and Astronomy, Texas A\& M University, College Station,
TX 77843, USA}

{email: {\tt sezgin@tamu.edu}}

\end{center}

\vskip 0.5cm

\begin{center} {\bf ABSTRACT}\\[3ex]
\end{center}

 There exists two distinct  off-shell $\cN=2$ supergravities in three dimensions. They are also referred to as $\cN=(1,1)$ and $\cN=(2,0)$ supergravities, and they arise from the coupling of the Weyl multiplet to a compensating scalar or vector multiplet, respectively, followed by fixing of conformal symmetries. The $\cN=(p,q)$ terminology refers to the underlying \mbox{anti-de Sitter} superalgebras $OSp(2,p) \oplus OSp(2,q)$ with $R$-symmetry group $SO(p) \times SO(q)$. We construct off-shell invariants of these theories up to fourth order in derivatives. As an application of these results, we determine the special combinations of the $\cN=(1,1)$ invariants that admit anti-de Sitter vacuum solution about which there is  a ghost-free massive spin-2 multiplet of propagating modes. We also show that the $\cN=(2,0)$ invariants do not allow such possibility.

\end{titlepage}

\newpage

\setcounter{page}{1}

\tableofcontents

%\newpage

%%%%%%%%%%%%%%%%%%%%%%%%%%%%%%%%%%%%
\section{Introduction}
%%%%%%%%%%%%%%%%%%%%%%%%%%%%%%%%%%%%

The time-honored motivation for studying three dimensional gravity theories is the prospects of their teaching us  lessons about much harder problem of gravity in four dimensions, at the classical as well as quantum level. That the black hole physics is nontrivial in three dimensions and one can extract valuable information from their study has been long recognized.  It is also well known that the important problem of massive gravity is much simpler to study and yet very rich in three dimensions. Furthermore, the quantization problem is more amenable as well in three dimensions though by no means trivial. In all these areas to explore, the role of higher derivative extensions is highly pertinent question.

The usual dictum that more symmetries give us more control over the theory motivates us to construct higher derivative supergravity theories with extended supersymmetry in three dimensions. In doing so, the role of off-shell versus on-shell nature of local supersymmetry brings in interesting new ingredients. Focusing our attention to supergravity theories which admit anti-de Sitter space as a vacuum solution, their underlying supersymmetry algebra is $OSp(p,q)$ whose bosonic part is  $O(2,2) \oplus SO(p) \times SO(q)$ \cite{Achucarro:1987vz,Achucarro:1989gm,Howe:1995zm}. Off-shell supergravity invariants up to and including four derivatives are known for $\cN=(1,0)$ supergravity in three dimensions \cite{Bergshoeff:2010mf}. A particular combinations of these invariants constitute the supersymmetric generalization of the so-called ``New Massive Gravity'' which has the virtue of being ghost-free \cite{Bergshoeff:2009hq}. Some of their properties, such as their supersymmetric vacua and spectrum about $AdS_3$ vacuum, have also been studied \cite{Bergshoeff:2010mf,Andringa:2009yc,Bergshoeff:2010iy}.

Our aim here is to generalize the construction of higher derivative supergravity invariants to those with underlying $\cN=(1,1)$ and $\cN=(2,0)$ superalgebras and to look for their ghost free combinations. The conformal $\cN=2$ supergravity, and the two-derivative invariants were considered in \cite{Rocek:1985bk,Nishino:1991sr, Cecotti:2010dg, Zupnik:1988en}. Off-shell matter-coupled suprgravity theories were investigated in the superspace framework in  \cite{Kuzenko:2013uya, Kuzenko:2011rd,Kuzenko:2014jra,Kuzenko:2011xg}. The on-shell construction and the matter couplings of the three dimensional $\cN=2$ supergravity were studied in \cite{Izquierdo:1994jz,de Wit:1992up,Deger:1999st}. Here we shall provide all the four derivative off-shell invariants of the $3D,~\cN=2$ supergravity. The method we shall employ is the superconformal tensor calculus. The $\cN=(1,1)$ and $\cN=(2,0)$ supergravities arise  from the coupling of Weyl multiplet to a compensating scalar or vector multiplets, respectively, followed by fixing of conformal symmetries.  In the case of $\cN=(2,0)$ supersymmetry, we shall employ a map between the Yang-Mills and the supergravity multiplet to construct the supersymmetric completion of the Ricci-squared term.

Taking into account the new invariants we construct here, we end up with seven parameter action with $\cN=(1,1) $ supersymmetry and a six parameter action with $\cN=(2,0)$ supersymmetry.  We find that the former admits a four parameter subfamily which admits AdS vacuum solution around which the spectrum of small fluctuations is ghost-free. In the latter case, however, we find that such a scenario is not possible. This turns out to be due to the fact that a particular type of dimension four invariant that exists for the $\cN=(1,1)$ model does not seem to exist for the $\cN=(2,0)$ model. The existence of the supersymmetric cosmological extension of the $(2,0)$ supergravity, which does not exist in the parent $\cN=1$ new minimal supergravity in $4D$, is not sufficient for the existence of a ghost-free supersymmetric $AdS_3$ vacuum.

This paper is organized as follows. In section \ref{section: 2}, we give a brief introduction to superconformal formalism, and introduce the Weyl multiplet, the scalar and the vector multiplet of the $D=3,~ \cN=2$ theory in the context of conformal supergravity. We then provide a multiplication rule for scalar multiplets, and construct composite scalar and vector multiplets. Subsequently, we proceed to construct various superconformal  actions for these matter multiplets. In section \ref{section: 3}, we consider the scalar multiplet actions constructed in section \ref{section: 2} and gauge fix the superconformal symmetries to obtain $\cN = (1,1)$ cosmological Poincar\'e supergravity as well as the supersymmetric completion of the $R_{\m\n}^2, R^2$ and the off-diagonal $(RS^n + h.c)$, where $S$ is the auxiliary scalar of the $\cN = (1,1)$ Poincar\'e multiplet. We then present the $\cN = (1,1)$ generalized massive supergravity, and analyze the bosonic spectrum around a maximally supersymmetric $AdS_3$ vacuum. In section \ref{section: 4}, we repeat the same analysis for the vector multiplet, and construct the $\cN = (2,0)$ cosmological Poincar\'e supergravity and the supersymmetric $R^2$ and off-diagonal $RD$ invariants, where $D$ is the auxiliary scalar of the $\cN = (2,0)$ Poincar\'e multiplet. This section contains an observation which states that certain fields in the $\cN =(2,0)$ Poincar\'e multiplet transform in the same manner as the fields of a Yang-Mills multiplet with $G = SO(2, 1)$. By use of that explicit correspondence, we construct the supersymmetric $R_{\m\n}^2$ invariant. Subsequently, we discuss the ghost-free $\cN = (2,0)$ generalized massive gravity, and analyze the spectrum around a maximally supersymmetric Minkowski background. In section \ref{section: 5}, we give conclusion and discussions. Finally, the details of the complex spinor conventions and Fierz identities are given in the appendix.

%%%%%%%%%%%%%%%%%%%%%%%
\section{Superconformal Tensor Calculus} \label{section: 2}
%%%%%%%%%%%%%%%%%%%%%%%

In this section we shall describe the Weyl multiplet based on the superconformal algebra $OSp(4|2)$ in three dimensions. We will then present the off-shell scalar and vector multiples which will be used in the subsequent sections as compensators. The rules for combining these multiplets to obtain new (composite) multiplets and action formula will follow. The  action formula will be used in the following sections together with the composite multiplet formula to obtain several off-shell supergravity invariants. Finally, we shall also record for completeness the Chern-Simons invariant which does not require any compensating multiplet coupling since it is superconformal invariant by itself \cite{Rocek:1985bk}.

%%%%%%%%%%%%%%%%%%%%%%%%%%%%%%%%
\subsection{The Weyl and Compensating Multiplets}
%%%%%%%%%%%%%%%%%%%%%%%%%%%%%%%%

\paragraph{Weyl multiplet.}
%%%%%%%%%%%%%%%%

The $\cN=2$ Weyl multiplet in three dimensions is based on the the conformal superalgebra $OSp(4|2)$ and consists of the fields
\be
(e_\mu{}^a, \psi_\mu, V_\mu, b_\mu, \omega_\mu{}^{ab}, f_\mu{}^a, \phi_\mu)\ ,
\ee
where $e_\mu{}^a$ is the dreibein, $\psi_\mu$ is the gravitino represented by a Dirac vector-spinor,  $V_\mu$ is the $U(1)$ R-symmetry gauge field, $b_\mu$ is the dilatation gauge field, $\omega_\mu{}^{ab}$ is the spin connection, $f_\mu{}^a$ is the conformal boost gauge field and $\phi_\mu$ is the special supersymmetry gauge field represented by a Dirac vector-spinor. The corresponding gauge parameters are
\be
(\xi^a, \epsilon, \Lambda, \Lambda_D, \Lambda^{ab}, \Lambda_K^a, \eta)\ .
\ee
The gauge fields $\omega_\mu{}^{ab}, \phi_\mu, f_\mu{}^a$ can be expressed in terms of the remaining fields by imposing the constraints \cite{Rocek:1985bk}
\be
{\widehat R}_{\mu\nu}^a (P)=0\ ,\qquad {\widehat R}_{\mu\nu}{}^{ab} (M)=0\ , \qquad {\widehat R}_{\mu\nu} (Q)=0\ ,
\label{con}
\ee
where the supercovariant curvatures associated with translations, Lorentz rotations and supersymmetry are defined as
\bea
{\widehat R}_{\mu\nu}{}^a (P) &=& 2(\partial_{[\mu}+b_{[\mu})\,e_{\nu]}{}^a +2\,\omega_{[\mu}{}^{ab}e_{\nu]b}
-\ft12(\bar{\psi}_{[\mu}\gamma^{a}\psi_{\nu]}+h.c.)\ ,
\nn\\
{\widehat R}_{\mu\nu}{}^{ab}(M) & = & 2\partial_{[\mu}\omega_{\nu]}{}^{ab}+2\,\omega_{[\mu}{}^{ac}\,\omega_{\nu]c}{}^{b}+8f_{[\mu}{}^{[a}e_{\nu]}{}^{b]}-\frac{1}{2}\bar{\psi}_{\mu}\gamma^{ab}\phi_{\nu}-\ft{1}{2}\bar{\phi}_{\mu}\gamma^{ab}\psi_{\nu}+h.c
\nn\\
{\widehat R}_{\mu v}(Q) & = & 2\partial_{[\mu}\psi_{\nu]}+\ft{1}{2}\omega_{[\mu}{}^{ab} \gamma_{ab}\,\psi_{\nu]}+b_{[\mu}\psi_{\nu]}-2\,\gamma_{[\mu}\phi_{\nu]}-2{\rmi V_{[\mu}\psi_{\nu]}}\ .
\eea
These constraints together with the Bianchi identity for ${\widehat R}_{\mu\nu} (P)$ also imply that the curvature associated with dilatation also vanishes, \emph {viz.} ${\widehat R}_{\mu\nu}(D)=0$. Solving the constraints~\eq{con} gives
\bea
{\omega}_\mu{}^{ab} & = & 2e^{\nu[a}\partial_{[\mu}e_{\nu]}{}^{b]}-e^{\nu[a}e^{b]\sigma}e_{\mu c}\,\partial_{\nu}e_{\sigma}{}^{c}+2e_{\mu}{}^{[a}b^{b]}+\ft{1}{2}\bar{\psi}_\mu\gamma^{[a}\psi^{b]}\nn\\
&& +\ft{1}{2}\bar{\psi}^{[a}\gamma^{b]}\psi_{\mu}+\ft{1}{2}\bar{\psi}^{[a}\gamma_{\mu}\psi^{b]}\ ,
\nn\\
\phi_\mu & = & -\gamma^{a}\widehat{R}_{\mu a}^{'}(Q)+\ft{1}{4}\gamma_{\mu}\gamma^{ab}\widehat{R}_{ab}^{'}(Q)\ ,
\nn\\
f_\mu^{a} & = & -\ft{1}{2}\widehat{R}'_{\mu}{}^{a}(M)+\ft{1}{8}e_{\mu}{}^{a}\widehat{R}^{'}(M)\ ,
\label{eq:dependent}
\eea
where  the prime in the curvatures used in \eqref{eq:dependent} means that the term including the field we are solving for is excluded. The transformation rules for the independent fields are given by
\bea
\delta e_{\mu}{}^{a} & = & -\Lambda^a{}_b \, e_{\mu}{}^{b}-\Lambda_{D} e_{\mu}{}^{a}+\ft{1}{2}\bar{\epsilon}\,\gamma^{a}\psi_{\mu}+h.c.\ ,
\nn\\
\delta\psi_{\mu} & = & -\ft{1}{4}\Lambda^{ab}\gamma_{ab}\psi_{\mu}-\ft{1}{2}\Lambda_{D}\psi_{\mu} +\mathcal{D_{\mu}\epsilon}-\gamma_{\mu}\eta+\rmi \Lambda\psi_{\mu}\ ,
\nn\\
\delta b_{\mu} & = & \partial_{\mu}\Lambda_{D}+2\Lambda_{K\mu}+\ft{1}{2}\bar{\epsilon}\,\phi_{\mu}-\ft{1}{2}\bar{\eta}\,\psi_{\mu}+h.c\ ,
\nn\\
\delta V_{\mu} & = & \partial_{\mu}\Lambda+\ft{1}{2} \rmi \bar{\epsilon}\,\phi_{\mu}+\ft{1}{2}\rmi \bar{\eta}\,\psi_{\mu}+h.c.\ ,
\label{weyl}
\eea
where
\be
\mathcal{D_{\mu}\epsilon}= \Big(\partial_\mu +\frac{1}{2}b_{\mu}+\frac14\omega_\mu{}^{ab}\gamma_{ab}- \rmi V_\mu \Big) \epsilon\ .
\ee
Finally, we give the transformation rule for $\phi_\mu$ for later convenience
\bea
\delta \phi_\mu &=& \cdots + \rmi \g^\n \widehat{F}_{\m\n} \e - \ft14 \rmi \g_\m \g \cdot \widehat{F}_{\m\n}\e  \ ,
\eea
where we have displayed the supercovariant terms and the ellipses refer to the remaining terms implied by the $OSp(4|2)$ algebra, 
and $\widehat{F}_{\m\n}$ is given by
\bea
\widehat{F}_{\m\n} &=& 2 \partial_{[\m} V_{\n]}  - \rmi \bar\psi_{[\m} \phi_{\n]} - \rmi \bar\phi_{[\m} \psi_{\n]} \ .
\eea

%%%%%%%%%%%%%%%%%%%%%%%
\paragraph{Scalar multiplet.}
%%%%%%%%%%%%%%%%%%%%%%%

The off-shell $\cN = 2 $ scalar multiplet with 4+4 degrees of freedom consists of a physical complex scalar $A$, a Dirac fermion $\chi$ and an auxiliary complex scalar~$\cF$ with the following transformation rules \footnote{See Appendix \ref{section: A} for the definition of $\widetilde \eta$ and the constant matrix $B$.}
\bea
\delta A & = & \frac{1}{2}\bar{\epsilon}\chi+ w \Lambda_{D}A- \rmi w \Lambda A\ ,
\nn\\
\delta\chi & = & \slashed{\mathcal D}A\,\epsilon-\frac{1}{2}\cF\left(B\epsilon\right)^{*}+2w A\,\eta
+(w+\ft12 )\Lambda_{D}\chi+\rmi (-w+1)\Lambda\chi\ ,
\nn\\
\delta \cF & = & -\tilde{\epsilon}\slashed{\mathcal{D}}\chi+2(w-\ft12)\tilde{\eta}\,\chi
+(w+1)\Lambda_{D}\cF+\rmi (-w+2)\Lambda\, \cF\ ,
\eea
where the supercovariant derivatives are given by
\bea
\mathcal{D}_\mu A & = & (\partial_{\mu} - w\,b_{\mu} + \rmi w V_{\mu})A-\ft{1}{2}\bar{\psi}_{\mu}\chi\ , \nn
\\
\mathcal{D}_\mu \chi & = & (\partial_{\mu} - (w+\ft12)b_{\mu} + \ft14 \,\o_{\m}{}^{ab}\, \g_{ab} + \rmi (w-1)V_{\mu} )\chi
-\slashed{\mathcal{D}}A\psi_{\mu} \nn\\
&& +\ft{1}{2}\cF(B\psi_{\mu})^{*}-2wA\phi_{\mu}\ .
\eea
Note that the lowest component has Weyl weight $w$ and $U(1)_R$ weight $-w$. Another multiplet with its lowest component having
Weyl weight $w$ and $U(1)_R$ weight $w$ can be obtained by charge conjugation
\bea
\delta A^{*} & = & \ft{1}{2}\tilde{\epsilon}\left(B\chi\right)^{*}+w\Lambda_{D}A^{*}+iw\Lambda A^{*}\ ,
\nn\\
\delta\left(B\chi\right)^{*} & = & \slashed{\mathcal{D}}A^{*}\left(B\epsilon\right)^{*}-\ft12F^{*}\epsilon
+2w A^{*}\left(B\eta\right)^{*}+(w+\ft12)\Lambda_{D}\left(B\chi\right)^{*} \nn\\
&& + \rmi \left(w-1\right)\Lambda\left(B\chi\right)^{*}\ ,
\nn\\
\delta \cF^{*} & = & -\bar{\epsilon}\slashed{\mathcal{D}}\left(B\chi\right)^*
+2(w-\ft12)\bar{\eta}(B\chi)^* +(w+1)\Lambda_D \cF^* \nn\\
&& + \rmi (w-2)\Lambda \,\cF^* \ ,
\eea
where the supercovariant derivatives are
\bea
\mathcal{D}_{\mu}A^{*} & = & (\partial_\mu -w b_{\mu}- \rmi w V_{\mu})A^{*}-\ft{1}{2}\tilde{\psi}_\mu (B\chi)^{*}\ ,
\nn \\
\mathcal{D}_{\mu}(B\chi)^{*} & = & \left(\partial_{\mu}-(w+\ft12) b_{\mu}  + \ft14 \o_{\m}{}^{ab} \g_{ab}
- \rmi (w-1) V_\m \right)(B\chi)^{*}
\nn \\
 &  & -\slashed{\mathcal{D}}A^{*}(B\psi_{\mu})^{*} +\ft{1}{2}\cF^{*}\psi_{\mu}-2w A^{*}(B\phi_\mu)^{*}\,,\nn\\
\cD_\m P^* &=& ( \partial_\m - \ft12 b_\m - (w - 2 ) \rmi \, V_\m ) P^* + \bar\p_\m \slashed{\cD} (B \chi)^* \nn\\
&& - 2 ( w - \ft12 ) \bar\phi_\m (B \chi)^* \,.
\eea
{\small
\begin{table}[h]
\begin{center}
\begin{tabular}{|c|c|c|c|c|c|}
\hline
Multiplet&Field& Type&Off-shell&$w$&q \\[.1truecm]
\hline\rule[-1mm]{0mm}{6mm}
Weyl&$e_\mu{}^a$&dreibein&2&-1 & 0\\[.1truecm]
&$\psi_\mu$&gravitino&4&$-\ft12$ & 1 \\[.1truecm]
&$V_\mu$&$U(1)_R$ gauge field&2&$0$ & 0 \\[.1truecm]
\hline
Scalar&$A$&complex scalar&2&$w_A$ & $- w_A$ \\[.1truecm]
&$\chi$&Dirac spinor&4&$w_A +\ft12$ &$ - w_A + 1 $\\[.1truecm]
&$\cF$&complex auxiliary&2&$w_A+1$ & $ -w_A + 2 $ \\[.1truecm]
\hline\rule[-1mm]{0mm}{6mm}
Vector&$\r$&real scalar&1&1& 0 \\[.1truecm]
&$C_\m$&gauge field&2&0 & 0 \\[.1truecm]
&$\lambda$&Dirac spinor&$4$&$\ft32$& 1\\[.1truecm]
&$D$&real auxiliary&1&2 & 0\\[.1truecm]
\hline
\end{tabular}
\end{center}
\caption{\footnotesize Properties of the $3D, \cN = 2$  Weyl and compensating multiplets where $(w,q)$ label the dilatation weight and  the $U(1)_R$ charge, respectively.}\label{table1}
\end{table}
}

{\small
\begin{table}[h]
\begin{center}
\begin{tabular}{|c|c|c|}
\hline
Components&$w$&$q$\\[.1truecm]
\hline\rule[-1mm]{0mm}{6mm}
$(\xi,\vf,M)$ &$ \ft52$& $- \ft52 $\\[.1truecm]
$(Z,\O,F)$ & 2& -2\\[.1truecm]
$(\phi,\zeta,S)$ &$\ft12$& $-\ft12$\\[.1truecm]
$(\sigma,\psi,N)$&$0$&$0$ \\[.1truecm]
$(\Phi, \Psi, P)$ & $-\ft12$ & $\ft12$\\[.1truecm]
\hline
\end{tabular}
\end{center}
\caption{\label{special}\footnotesize  Compensating scalar multiplets  with $(w,q)$ denoting the Weyl weight and
$U(1)_R$ charge of the lowest component scalar field.}
\label{Table2}
\end{table}
 }

%%%%%%%%%%%%%%%%%%
\paragraph{Vector multiplet.}
%%%%%%%%%%%%%%%%%%

The off-shell vector $\cN=2$ vector multiplet with~$4+4$ degrees of freedom consists of a gauge field~$C_{\mu}$, a scalar~$\rho$,
a spinor~$\lambda$ and an auxiliary scalar~$D$. Their transformation rules are given by
\bea
 \delta C_{\mu} &=& \ft{1}{2}\,\bar{\epsilon}\, \gamma_{\mu} \lambda - \ft{1}{4} \rmi\, \rho\, \bar{\epsilon}\, \psi_{\mu} + h.c. \ ,
 \nn\\
 \delta \rho &=&  (\rmi \bar{\epsilon}\, {\l} + h.c. )+ \Lambda_D \rho \ ,
 \nn\\
 \delta \lambda &=&  - \ft{1}{4} \gamma^{\mu\nu} \widehat{G}_{\mu\nu}\, \epsilon +  \ft{1}{2} \rmi D \epsilon
 - \ft{1}{4}  \rmi \slashed{\mathcal{D}} \rho \,\epsilon -  \ft{1}{2} \rmi \rho \,\eta + \rmi \Lambda\, \lambda +\ft{3}{2} \Lambda_D \lambda\ ,
 \nn\\
 \delta D &=& ( - \ft{1}{2} \rmi \bar{\epsilon} \slashed{\mathcal{D}} \lambda+ \ft{1}{2} \rmi \bar{\eta} \lambda+ h.c. ) +2 \Lambda_D D\ ,
 \label{vtr}
\eea
where
\bea
 \mathcal{D}_{\mu}\rho & = & \left(\partial_{\mu} - b_{\mu}\right) \rho + \left(-\rmi \bar{\psi}_{\mu}\lambda + h.c. \right)\ ,
 \nn\\
 \mathcal{D}_{\mu}\lambda & = & (\partial_{\mu} - \ft{3}{2} b_{\mu} +\ft 14\,\omega_\mu{}^{ab}\gamma_{ab}
 - \rmi V_{\mu}) \lambda
 + \ft{1}{4} \gamma^{\rho\sigma} \widehat{G}_{\rho\sigma}\, \psi_{\mu}
 \nn\\
&&  - \ft{1}{2} \rmi D\psi_{\mu}
 + \ft{1}{4} \rmi \slashed{\mathcal{D}} \rho\, \psi_{\mu} + \ft{1}{2}\rmi \rho\, \phi_{\mu}\ ,
 \nn\\
 \widehat{G}_{\mu\nu} & = & 2 \partial_{[\mu} C_{\nu]} + (-\bar{\psi}_{[\mu} \gamma_{\nu]} \lambda
 + \ft{1}{2}\rmi \rho\, \bar{\psi}_{\mu} \psi_{\nu} + h.c.)\ .
 \label{vd}
\eea
As we shall discuss in the subsection~\ref{ss: 43}, the nonabelian versions of \eq{vtr} and \eq{vd} can be obtained by taking the fields of the vector multiplet in adjoint representation of a Lie group $G$, and imposing the closure of the algebra accordingly.

%%%%%%%%%%%%%%%%%%%%%%%%%%%%%%%%%%%%%
\subsection{Combination of Local Supermultiplets}
%%%%%%%%%%%%%%%%%%%%%%%%%%%%%%%%%%%%%

To provide a supersymmetric completion of the Poincar\'e supergravity and of the higher dimensional invariants, we need to produce multiplets with different weights. We thus give now general rules to do so and we introduce all  composite multiplets that will be needed to construct invariant actions.

\paragraph{Scalar Multiplets.}
%%%%%%%%%%%%%%%%%%%%%%%

We will construct  composite scalar multiplets using the multiplication rules for scalar multiplets. One can start with two scalar multiplets
$\left(A_{i},\chi_{i},F_{i}\right)$, $i=1,2$ and obtain a multiplet whose lowest component having Weyl weight $w=w_1+w_2$
and $U(1)_R$ weight $q=q_{1}+q_{2}$ as follows
\bea
A & = & A_{1}A_{2}\ ,
\nn\\
\chi & = & A_{1}\chi_{2}+A_{2}\chi_{1},\nonumber \\
F & = & A_{1}F_{2}+A_{2}F_{1}+\tilde{\chi}_{1}\chi_{2}\ .
\label{m1}
\eea
It is also possible to use the inverse of the multiplication rule \eqref{m1} to obtain a multiplet with Weyl weight
$w=w_1-w_2$ and $U(1)_R$ weight $q=q_{1}-q_{2}$
\bea
A & = & A_{1}A_{2}^{-1}\ ,
\nn\\
\chi & = & A_{2}^{-1}\chi_{1}-A_{1}A_{2}^{-2}\chi_{2}\ ,
\nn\\
F & = & A_{2}^{-1}F_{1}-A_{1}A_{2}^{-2}F_{2}-A_{2}^{-2}\tilde{\chi}_{2}\chi_{1}+A_{1}A_{2}^{-3}\tilde{\chi}_{2}\chi_{2}\ .
\label{eq:inversemult}
\eea
Given the scalar multiplet (see {table~\ref{table1}),
\be
\Sigma = \left(\phi,\zeta,S\right)\ ,
\ee
the associated inverse multiplet has the components
\be
\Sigma^{-1}  \equiv (\Phi,\Psi,P) = \left(\phi^{-1}\ , \ \  -\phi^{-2}\zeta, \ \ -\phi^{-2}S+\phi^{-3} {\tilde\zeta} \zeta\right) \ .
\label{com3}
\ee
as can be seen by considering the multiplication of the unit multiplet $(A_1, \chi_1, F_1)=(1,0,0)$, which has weights $(\omega,c)=(0,0)$, with the multiplet $(A_2, \chi_2, F_2)= \Sigma$,  by means of the formula~\eq{eq:inversemult}.

Next, we note that a scalar multiplet $\left(\phi,\zeta,S\right)$ with weights $(w,q) = (\frac12, -\frac12)$ has the
corresponding kinetic multiplet with weights $(w,q)= (\frac{3}{2}, - \frac{3}{2})$ given by
\be
 {\cal K} = \left( S^*, -2\slashed{\mathcal{D}}(B\zeta)^*, 4\Box^{C}\phi^* \right)\ ,
\label{kin}
\ee
where
\bea
\Box^{C}\phi^{*} &=& \left(\partial^{a}-\ft{3}{2}b^{a}-\ft{i}{2}V^{a}\right)\mathcal{D}_{a}\phi^{*}
+\omega_{a}{}^{ab}\mathcal{D}_{b}\phi^{*}+f_{a}^{a}\phi^{*}\nn\\
 && +\ft{1}{2}\tilde{\phi}_{a}\gamma^{a}(B\zeta)^{*}-\ft{1}{2}\tilde{\psi}^{a}\mathcal{D}_{a}(B\zeta)^{*}\ .
\eea
Using the above multiplets as building blocks and using the product formula \eq{m1} we can construct a number of multiplets which will be useful in building actions. To begin with, we consider the four-fold product of $\Sigma$ obtaining
\be
\Sigma^4:\qquad (Z,\Omega,F)= \left(\phi^{4},\ \ 4\phi^{3}\zeta,\ \ 4\phi^{3}S+6\phi^{2}{\tilde \zeta}\zeta\right) \ .
\label{com1}
\ee
Note that $Z$ has the weights $(w,q)=(2,-2)$ and will be useful to construct a cosmological constant invariant. Another multiplet with the same weights~$(2, -2)$ is
\bea
\Sigma \times {\cal K} \quad :\quad
Z '&=& \phi S^*\ ,
\nn\\
\Omega' &=& \zeta S^* -2\phi \slashed{\cD}(B\zeta)\ ,
\nn\\
F' &=& 4\phi \Box^C\phi^* + |S|^2  -2{\tilde\zeta} \slashed{\cD} (B\zeta)^*\ .
\label{com2}
\eea
We will use this multiplet to construct the Einstein-Hilbert action. A composite neutral multiplet with~$(w,q) = (0,0)$ can be obtained as follows
\bea
 {\cal K} \times \Sigma^{-3}&:&
\nn\\
 \sigma &=& \phi^{-1} S^*\ ,
 \nn\\
 \psi&=& -2\phi^{-3}\slashed{\mathcal{D}}\left(B\zeta\right)^{*}-3\phi^{-4}S^{*}\zeta\ ,
\nn\\
N & = & 4\phi^{-3}\Box^{C}\phi-3\phi^{-4}\left|S\right|^{2}+6\phi^{-4}\tilde{\zeta}\slashed{\mathcal{D}}\left(B\zeta\right)^{*}
+6\phi^{-5}S^{*}\tilde{\lambda}\zeta\ ,
\label{com4}
\eea
which can be used to produce new scalar multiplets without changing the weights of the original multiplets
\bea
(\sigma,\psi,N)^n \times (Z,\Omega,F) &:&
\nn\\
Z^{(n)} &=& \sigma^n Z\ ,
\nn\\
\Omega^{(n)} &=& n \s^{n-1} Z \p + \s^n \O\ ,
\nn\\
F^{(n)} &=&  \sigma^n F + n \sigma^{n-1} Z N + n (n-1) \s^{n-2} Z \tilde\p \p \nn\\
&& + n \s^{n-1} \tilde\p \O \ .
\label{com6}
\eea
\be
(\sigma,\psi,N) \times \Sigma:\qquad (\phi',\zeta',S') = \left( \sigma\phi\ ,\ \ \sigma\zeta +\phi\psi\ ,\ \ \sigma S+\phi N+{\tilde\zeta}\psi\right) \ .
\label{com5}
\ee
Finally, we construct the multiplet $(\xi,\vf,M)$, with weights~$(\frac{5}{2}, -\frac{5}{2})$, in terms of the elements of the multiplet $(\Phi, \Psi, P)$ as }
\bea
\xi &=& \Box^c P^*  \ ,
\nn\\
\vf &=& -2 \Box^c \slashed{\cD} (B \Psi)^*  -2 \rmi \g^\n \cD^\m \widehat{F}_{\m\n} (B\l)^* + 2 \rmi \g^\n  \widehat{F}_{\m\n} \cD^\n (B\l)^*
\nn\\
&& + \rmi \g^{\m\n} \slashed{\cD}  \widehat{F}_{\m\n} (B \l)^* + \ft52 \rmi \g^{\mu\nu}  \widehat{F}_{\mu\nu} \slashed{\cD} (B\l)^* \,,\nn\\
M &=& 4 \Box^c \Box^c \Phi^* - 8 \rmi \cD^a \widehat{F}_{ab}  {\mathcal{D}^b}\Phi^{*} - 2 \widehat{F}_{ab} \widehat{F}^{ab} \Phi^* + \text{fermions} \ ,
\label{eq:mudoubox}
\eea
where we have omitted the complicated fermionic expressions in the composite formula for~$M$ as we shall be interested in the bosonic part of an action formula for which this multiplet will be used. With this multiplet we will produce a Ricci tensor squared invariant.

%%%%%%%%%%%%%%%%%%%%%%%
\paragraph{Vector Multiplets.}
%%%%%%%%%%%%%%%%%%%%%%%
For the construction of the $n$ vector multiplet action, we first introduce a real function $C_{IJ} (\rho)$, which is a function of the vector multiplet scalars $\r^I$, and the $n$ vector multiplets are labeled by $I,J,\ldots = 1,2,\ldots , n$. The lowest component of a vector multiplet can then be composed as
\bea
\rho_I  & = & C_{IJ}D^{J} +C_{IJK}\,\bar{\l}^{J}\l^{K}\ .
\label{rho}
\eea
{The label $I$ is fixed, and it differs from the indices that are being summed over. We also define
\bea
 C_{IJK}=\frac{\partial C_{IJ}}{\partial\rho^{K}}\ ,\quad C_{IJKL}
=\frac{\partial^{2}C_{IJ}}{\partial\rho^{K}\,\partial\rho^{L}} \ ,\qquad 
C_{IJKLM}
=\frac{\partial^{3}C_{IJ}}{\partial\rho^{K}\,\partial\rho^{L}\,\partial\rho^{M}} \ .
\eea
In order to ensure that the $\r_I$ is the scalar of a superconformal vector multiplet, we impose that the conformal weight of
$C_{IJ}$ is $\o (C_{IJ}) = -1$, and following constraints are satisfied
\bea
C_{IJK}=C_{I(JK)}\ , \quad C_{IJK}\, \r^K =  - C_{IJ} \ .
\label{CIJConstraints1}
\eea
Furthermore, additional constraints are needed to ensure that $\l_I, D_I$ and $\widehat{G}_{\m\n I}$ are also the elements of a superconformal vector multiplet
\bea
 C_{IJKL}\, \r^L = - 2\, C_{IJK}\,, \quad C_{IJKLM}\, \r^M = -3\, C_{IJKL}\ .
\eea
Therefore, applying a sequence of Q- and S-transformations, we find the elements of the composite vector multiplet as
\bea
\rho_I  & = & C_{IJ}\,D^{J} +C_{IJK}\,\bar{\l}^{J}\l^{K}\ ,
\nn\\
\l{}_{I} & = & \ft12\, C_{IJK}\,D^{J}\l^{K}-\ft12\, C_{IJ}\,\slashed{\mathcal{D}}\l^{J}
-\ft{1}{4} \rmi\, C_{IJK}\,\gamma^{\mu\nu}\widehat{G}_{\mu\nu}^{J}\l^{K}
\nn\\
 & & -\ft14\, C_{IJK}\,\slashed{\mathcal{D}}\rho^{J}\l^{K}+C_{IJKL}\,\l^{L}\bar{\l}^{J}\l^{K}
 \nn\\
D_I & = & \ft12 \,C_{IJK}\,D^{J}D^{K}+\ft{1}{4}\,C_{IJ}\,\Box^{C}\rho^{J} - \ft14 \,C_{IJK}\,\widehat{G}_{\mu\nu}^{J}\,\widehat{G}^{\mu\nu K}
\nn \\
 & & +\ft18 \,C_{IJK}\,\mathcal{D}_{\mu}\rho^{J}\mathcal{D}^{\mu}\rho^{K} - \ft12 \,C_{IJK}\, \bar\l^J \slashed{\cD} \l^K + \ft12 \,C_{IJK} \,\overline{\cD_\m \l^J} \g^\m  \l^K \ ,
\nn\\
&& - \ft12 \rmi \,C_{IJKL} \,\bar\l^L \g^{\mu\nu} \widehat{G}_{\mu\nu}^J \l^K + C_{IJKL}\, D^J \bar\l^K \l^L + C_{IJKLM} \,\bar\l^J\l^K\bar\l^L\l^M \ ,
 \nn\\
\widehat{G}_{\mu\nu I} & = & \ft12 \mathcal{D}_{\sigma}\left(\epsilon_{\lambda\mu\nu}\,C_{IJ}\,\widehat{G}^{\sigma\lambda J}\right) + 2 \rmi  \cD_{[\m}  \Big( C_{IJK} \,\bar\l^J \g_{\n]} \l^K \Big) -\ft14 \,C_{IJ}\,\rho^{J}\widehat{F}_{\mu\nu} \ ,
\label{compVV}
\eea
where the superconformal d'Alambertian for $\rho^{I}$ is given by
\be
\Box^{C}\rho^{I}=\left(\partial^{a}-2b^{a}+\o_{b}{}^{ba}\right)\mathcal{D}_{a}\rho^{I}+2f_{a}^{a}\rho^{I}+\left(-\rmi \bar{\psi}^{a}\mathcal{D}_{a}\l^{I}+i\bar{\phi}_{a}\gamma^{a}\l^{I}+h.c.\right)\,.
\ee
Note that $\widehat{G}_{\m\nu I}$ satisfies the Bianchi identity due to the constraints (\ref{CIJConstraints1}).

The composition formula (\ref{compVV}) can be truncated to a map between two vector multiplets
by choosing $C_{21}=\rho^{-1}$, in which case one obtains, for the bosonic fields,
\bea
\rho^{\prime} & = & \rho^{-1}D-\rho^{-2}\bar{\lambda}\lambda\ ,
\nonumber \\
D^{\prime} & = & -\ft{1}{2}\rho^{-2}D^{2}+\frac{1}{4}\rho^{-1}\Box^{C}\rho + \ft{1}{4}\rho^{-2}\widehat{G}_{\mu\nu}\,\widehat{G}^{\mu\nu} -\ft{1}{8}\rho^{-2}\mathcal{D}_{\mu}\rho\,\mathcal{D}^{\mu}\rho
 \nonumber \\
\widehat{G}'_{\mu\nu} & = & \ft12 \mathcal{D}_{\sigma}\left(\epsilon_{\lambda\mu\nu}\rho^{-1}\widehat{G}^{\sigma\lambda}\right)
-\ft{1}{4}\widehat{F}_{\mu\nu} \ ,
\label{nc1}
\eea
where $1$ labels the multiplet $(\r, C_\m, \l, D)$, and $2$ labels the multiplet $(\r', C_\m', \l', D')$. Another composite multiplet is obtained by choosing $C_{31}=-\rho^{-2}\rho^{\prime}$ and $C_{32}=\rho^{-1}$ in the composition formula \eq{compVV}. The bosonic components of the composite multiplet  $(\rho^{\prime\prime},\lambda^{\prime\prime},C_{\mu}^{\prime\prime},D^{\prime\prime})$, labeled by 3,  can then be written as
\bea
\rho^{\prime\prime} & = & -\rho^{-2}\rho^{\prime}D+\rho^{-1}D^{\prime}\ ,
\nonumber \\
D^{\prime\prime} & = & \rho^{-3}\rho^{\prime}D^{2}-\rho^{-2}DD^{\prime}-\ft14 \rho^{-2}\rho^{\prime}\Box^{C}\rho
+\ft14\rho^{-1}\Box^{C}\rho^{\prime}
\nonumber \\
 && - \ft12 \rho^{-3}\rho^{\prime}\widehat{G}_{\mu\nu}\,\widehat{G}^{\mu\nu} + \ft12 \rho^{-2}\widehat{G}_{\mu\nu}^{\prime}\widehat{G}^{\mu\nu}
 +\ft14 \rho^{-3}\rho^{\prime}\mathcal{D}_{\mu}\rho\,\mathcal{D}^{\mu}
 \rho\nonumber \\
 && -\frac{1}{4}\rho^{-2}\mathcal{D}_{\mu}\rho^{\prime}\mathcal{D}^{\mu}\rho\ ,
 \nonumber \\
\hat{G}_{\mu\nu}^{\prime\prime} & = & \ft12 \epsilon_{\lambda\mu\nu}\mathcal{D}_{\sigma}\left(-\rho^{-2}\rho^{\prime}\widehat{G}^{\sigma\lambda}
+\rho^{-1}\widehat{G}^{\prime\sigma\lambda}\right)\ .
\label{nc2}
\eea
%

%%%%%%%%%%%%%%%%%%%%%%%%%%%%%%%%%%%
\subsection{Action Formulae}
%%%%%%%%%%%%%%%%%%%%%%%%%%%%%%%%%%%

In this section, we collect the action formulae for scalar and vector multiplets that we shall use in the subsequent sections when constructing supergravity models. The construction of the supergravity invariants require coupling the Weyl multiplet to at least one compensating multiplet. We, therefore, consider two classes of supersymmetric invariants in accordance with the compensating multiplet under consideration. Finally, we also present the superconformal completion of the Chern-Simons action \cite{Rocek:1985bk}, which does not require a compensating multiplet.

\paragraph{Scalar Multiplet Actions.}

We start with the action for a scalar multiplet $\left(Z,\Omega,F\right)$
\be
e^{-1}\mathcal{L}_F = Re\left(F-\tilde{\psi}_{\mu}\gamma^{\mu}\Omega-Z\tilde{\psi}_{\mu}\gamma^{\mu\nu}\psi_{\nu}\right)\ ,
\label{F}
\ee
which is invariant under dilatations and $U(1)_R$ transformations since the highest component field $F$ has the weight $(w,q)= (3,0)$.

The composite multiplet given in \eq{com6} can be used in  the action formula \eq{F} to produce
\bea
e^{-1} \cL_{F^{(n)}} &=&  {\rm{Re}} \Big( \sigma^n F + n \,\sigma^{n-1} Z N + n (n-1) \s^{n-2} Z \tilde\p \, \p + n\, \s^{n-1} \tilde\p \O 
\nn\\
&&\qquad -n \s^{n-1} Z \,\tilde\p_\m \g^\m  \p - \s^n \tilde\p_\m \g^\m \, \O - \s^n Z\, \tilde\p_\m \g^{\m\n} \p_\n \Big) \ ,
\label{Ln}
\eea
which we shall use below to obtain an action providing a supersymmetric completion of~$RS^n$.

Next, we use the components of the composite scalar multiplet  multiplet \eq{com2} in the action formula \eq{F} which yields  the following action that will be used to construct the supersymmetric completions of the Einstein-Hilbert term as well as the $R^2$ term
\bea
e^{-1} {\mathcal L}_{\rm K} &=& 4\phi\,\Box^{C}\phi^{*}+\left|S\right|^{2}-2\,\tilde{\zeta}\,\slashed{\mathcal{D}}\left(B\zeta\right)^{*}+2\,\phi\,\tilde{\psi}_{\mu}\gamma^{\mu}\slashed{\mathcal{D}}\left(B\zeta\right)^{*}
\nn\\
&& -S^{*}\tilde{\psi}_{\mu}\gamma^{\mu}\zeta-\phi \,S^{*}\tilde{\psi}_{\mu}\gamma^{\mu\nu}\psi_{\nu}\ .
\label{K}
\eea
We next consider the scalar multiplets $(\xi,\vf,M)$ and $(\Phi, \Psi, P)$. Using the multiplication rule (\ref{m1}), the action describing the coupling of these multiplets can be given by
\bea
e^{-1} \cL_{\xi \Phi} &=& {\rm{Re}} \Big( \xi P + \Phi M  + \tilde{\Psi} \vf - \Phi \,\tilde{\psi}_\m \g^\m \vf - \xi \tilde{\p}_\m \g^\m \Psi - \Phi\, \xi \,\tilde{\p}_\m \g^{\m\n} \p_\n \Big) \ .
\label{RmnInt}
\eea
 Using the composite expressions (\ref{eq:mudoubox}), the bosonic part of the action that gives rise to the~$R_{\mu\nu}^2$ invariant is given by
\be
e^{-1} \mathcal{L}_{\Phi}= {\rm{Re}} \Big( 4\Phi\,\Box^{C}\Box^{C}\Phi^{*}+P\Box^{C}P^{*}   
- 8 \rmi \Phi\, \cD^a \widehat{F}_{ab}  {\mathcal{D}^b}\Phi^{*} - 2 \widehat{F}_{ab}\, \widehat{F}^{ab}\Phi \Phi^* \Big) \ .
\label{Phi}
\ee

\paragraph{Vector Multiplet Actions.}

Supersymmetric Lagrangians for vector multiplet can be constructed starting from an action formula which describes coupling of two vector multiplets as
\bea
e^{-1} \mathcal{L}_{D D^\prime}  &=& \rho D' + \rho' D + 2 \left( {\bar\lambda} \lambda' + h.c. \right) - 2\epsilon^{\mu\nu\rho} C_\mu \partial_\nu C'_\rho
\nn\\
&& - \ft{1}{2} \rmi \left( \rho\, {\bar\psi}_\mu \gamma^{\mu} \lambda' + \rho' {\bar\psi}_\mu\gamma^{\mu} \lambda+ h.c. \right)
 - \ft{1}{8} \left( \rho \rho' {\bar\psi}_\mu  \gamma^{\mu\nu} \psi_\nu + h.c. \right) \ .
\label{D}
\eea
As a special case, one can set the primed and the un-primed multiplet equal to each other, obtaining \cite{Kuzenko:2014jra}
\bea
e^{-1} \cL_{D} &=& 2  \rho D - \e^{\m\n\r} C_\m\, G_{\n\r} + 4 \bar\l \,\l
- \rmi  \left( \rho\, {\bar\psi}_\mu \gamma^{\mu} \lambda + h.c. \right) 
\nn\\
&&-\ft14 (\rho^2 {\bar\psi}_\mu  \gamma^{\mu\nu} \psi_\nu + h.c )\ .
\label{Dprime}
\eea

Using the composite multiplets \eq{nc1} in this action formula, we also obtain the conformal vector multiplet action
\bea
e^{-1}\mathcal{L}_{V} & = & \ft{1}{4}\Box^{C}\rho+\ft{1}{2}\rho^{-1}D^{2}
-\ft{1}{8}\rho^{-1}\partial_{\mu}\rho\,\partial^{\mu}\rho - \ft{1}{4}\rho^{-1}G_{\mu\nu}\,G^{\mu\nu} 
+\ft{1}{2}\epsilon^{\mu\nu\rho}\,C_{\mu}\,\partial_{\nu}V_{\rho}\ ,
 \label{eq:ConfVec}
\eea
up to fermionic terms.

Considering the coupling of a primed and double-primed multiplets in accordance with the action formula (\ref{D}), and employing the composite expressions \eq{nc2} result in an action that will be used in the construction of a supersymmetric completion of $R^2$ term,
\bea
e^{-1}\mathcal{L}_{VV'} & = &\r^{-3} (\rho^{\prime})^{2}D^{2} -2 \r^{-2} \rho^{\prime}DD^{\prime} + \r^{-1}(D^{\prime})^2 +\ft14\r^{-1} \rho^{\prime}\Box^c \rho^{\prime}
\nonumber \\
 &  & - \ft14 \r^{-2} \r^{\prime 2} \Box^c \r + \ft14 \r^{-3} \r^{\prime 2} \cD_\m \r\, \cD^\m \r - \ft14 \r^\prime \r^{-2} \cD_\m \r^\prime \cD^\m \r \nn\\
&& - \ft12 \r^{-3} (\rho^{\prime})^{2}\widehat{G}_{\mu\nu}\,\widehat{G}^{\mu\nu} + \rho^{\prime} \r^{-2} \widehat{G}_{\mu\nu}^{\prime}\,\widehat{G}^{\mu\nu} - \ft12 \r^{-1} \widehat{G}_{\mu\nu}^{\prime}\,\widehat{G}^{\prime\mu\nu}\,,
\label{PreYM}
\eea
where we have provided the terms that contributes to the bosonic part of the action. More generally, we obtain the most general $2-$derivative vector multiplet coupling, by using the action formula \eq{D}, as
\bea
e^{-1} \cL_{V_I} &=&  \ft14 C_{IJ}\, \r^I \Box^c \r^J + \ft18 C_{IJK}\, \r^I \cD_\m \r^J \cD^\m \r^K    - \ft12 C_{IJ}\, \widehat{G}_{\m\n}^I \widehat{G}^{\m\n J}
\nn\\
&& - \ft14 C_{IJK} \,\r^I \widehat{G}_{\m\n}^J \widehat{G}^{\m\n K}   + C_{IJ}\, D^I D^J + \ft12 C_{IJK}\, \r^I  D^J D^K \nn\\
&& + \ft14 C_{IJ}\, \r^J \e^{\m\n\r} C_\m^I\, F_{\n\r}\,.
\label{eq:VecGen}
\eea
Note that the index $I$ is fixed to represent a certain multiplet by construction due to~\eq{compVV}, and summing over $I$ indices correspond to summing different off-shell invariants.

Finally, there also exists an action that constitutes the superconformal completion of the Lorentz Chern-Simons term. It is given by \cite{Rocek:1985bk}
\be
{\cal L}_{\rm CS}  =  -\ft14\,\varepsilon^{\mu\nu\rho}\left[ R_{\mu\nu}{}^{ab}(\omega)\, \omega_{\rho ab}
+ \ft23 \omega_\mu{}^{ab}\, \omega_{\nu b}{}^c\,\omega_{\rho ca} \right]
 +\varepsilon^{\mu\nu\rho} F_{\mu\nu} {V_\rho} - \,{\bar R}^\mu \,\gamma_\nu\gamma_\mu R^\nu\ ,
\label{CS1}
\ee
where the Hodge dual of the gravitino curvature is defined by
\be
R^\mu  = \varepsilon^{\mu\nu\rho} (D_\nu(\omega)- \rmi {V_\nu}) \psi_\rho\ .
\ee
The supersymmetric Chern-Simons action is invariant under the Weyl multiplet transformation rules (\ref{weyl}). Therefore, it can be used for both $\cN =(1,1)$ and $\cN = (2,0)$ supergravities.
%We provide an alternative derivation of this Lagrangian in appendix C by using a map between Yang-Mills and %$\cN=(2,0)$ multiplet.

%%%%%%%%%%%%%%%%%%%%%%%%%%%%%%%%%%%
\section{${\cN}=(1,1)$ Supergravity Models} \label{section: 3}
%%%%%%%%%%%%%%%%%%%%%%%%%%%%%%%%%%%

The off-shell $\cN=(1,1)$ Poincar\'e supergravity and the supersymmetric completion of the cosmological term are already given in the literature, and they are also referred to as Type I minimal supergravity or three dimensional old minimal supergravity \cite{Rocek:1985bk, Nishino:1991sr, Cecotti:2010dg, Kuzenko:2013uya}. Here we shall derive them from the superconformal tensor calculus point of view, which will also serve to establish our notation and conventions. Again using the superconformal tensor calculus we shall construct three new invariants, namely the supersymmetric completion of the $R^2$ and $R_{\mu\nu}^2$ terms and of $(R S^2 + {\rm h.c.})$, where $S$ is the complex auxiliary field. The last invariant is key to the construction of ghost-free massive supergravity with  $\cN=(1,1)$ supersymmetry.

%%%%%%%%%%%%%%%%%%%%%%%%%%%%%%%%%%%%%%
\subsection{$\cN=(1,1)$ Cosmological Poincar\'e Supergravity}
%%%%%%%%%%%%%%%%%%%%%%%%%%%%%%%%%%%%%%

The off-shell Poincar\'e supergravity action is readily obtained from the action formula \eq{K} by fixing the dilatation, conformal boost and special supersymmetry transformation by imposing
\bea
\phi =1\ , \qquad   \zeta=0\ ,    \qquad b_\mu=0 \ .
\label{gf}
\eea
The first one fixes dilatation and $U(1)_R$ transformation, the second fixes the $S$-supersymmetry and the last one fixes the special conformal
transformations. Maintaining these gauge conditions imply that
\bea
\Lambda_{D} & = & \rmi \Lambda=0\ ,
\nn\\
\Lambda_{K\mu} & = & \frac{1}{4}\bar{\eta}\,\psi_{\mu}-\frac{1}{4}\bar{\epsilon}\,\phi_{\mu}+h.c. \ ,
\nn\\
\eta & = & -\frac{1}{2} \rmi \gamma^{\nu}V_{\nu}\epsilon+\frac{1}{2}S\left(B\epsilon\right)^{*}\ .
\label{gp}
\eea
These imply the super supersymmetry transformation rules
\bea
\delta e_\mu{}^{a}  &= & \ft{1}{2}\bar{\epsilon}\gamma^{a}\psi_{\mu}+h.c.
\nn\\
\delta\psi_\mu  &= & D_{\mu}(\omega)\,\epsilon-\ft{1}{2} \rmi V_{\nu}\,\gamma^{\nu}\gamma_{\mu}\,\epsilon
-\ft12 S\gamma_\mu \left(B\epsilon\right)^{*}
\nn\\
\delta V_\mu  &= & \ft{1}{8} \rmi \bar{\epsilon}\,\gamma^{\nu\rho}\gamma_{\mu}\left(\psi_{\nu\rho}-{\rmi} V_{\sigma}\gamma^{\sigma}\gamma_{\nu}\,\psi_{\rho}-S\gamma_\n \left(B\psi_{\rho}\right)^{*}\right)+h.c.
\nn\\
\delta S  & = & -\ft14 \tilde{\epsilon}\,\gamma^{\mu\nu}\left(\psi_{\mu\nu}
-\rmi V_{\s}\,\gamma^{\s}\gamma_{\mu}\psi_{\nu}-S\gamma_{\mu}\left(B\psi_{\nu}\right)^{*}\right)\ ,
\label{susy1}
\eea
where
\be
D_{\mu}(\omega)\epsilon=(\partial_{\mu}+\ft14\omega_{\mu}{}^{ab}\,\gamma_{ab})\epsilon\ ,
\qquad\psi_{\mu\nu}=2D_{[\mu}(\omega)\psi_{\nu]}\ .
\ee
Using the gauge fixing conditions \eqref{gf} in the action \eqref{K} gives
the the action of Poincar\'e supergravity
\be
e^{-1}\mathcal{L}_{EH} =  R+2V^2-2\left|S\right|^{2}
                                        -\left(\bar{\psi}_{\mu}\,\gamma^{\mu\nu\rho} D_{\nu}(\omega)\psi_{\rho}+h.c.\right)\ .
\label{EH}
\ee
where $V^2 := V_\mu V^\mu$. Next, we construct the supersymmetric cosmological term by using the multiplet $(Z,\O,F)$  given in \eq{com1} in the action formula \eq{F}, imposing the gauge fixing conditions \eqref{gf}, and multiplying the action by~$1/2$, arriving at the result
\be
e^{-1}\mathcal{L}_{C}=S-\frac{1}{4}\tilde{\psi}_{\mu}\,\gamma^{\mu\nu}\,\psi_{\nu}+h.c.
\label{eq:cosmological}
\ee

%%%%%%%%%%%%%%%%%%%%%%%%%%%%%%%%
\subsection{$\cN=(1,1)$ Higher Dimensional  Invariants}
%%%%%%%%%%%%%%%%%%%%%%%%%%%%%%%%

We begin with the construction of the $R^2$ invariant. To this end, we employ the composite scalar multiplet $\Sigma'$ from \eq{com5} in the action formula \eq{K}. In the resulting action we use the composite neutral multiplet  from \eq{com4}. Subsequently we fix the extra gauge symmetries as in \eq{gf}. These are straightforward manipulations which give the full $R^2$ invariant whose bosonic part is given by
\bea
e^{-1}{\mathcal L}_{R^2} &= & R^{2}+16\left|S\right|^{4}+ 4(V^2)^2 +6 R\left|S\right|^{2}+4RV^2+12\left|S\right|^2 V^2
\nn\\
&&  -16\partial_\mu S\,\partial^\mu S^* -8\rmi V^\mu S^* \overleftrightarrow{\partial_\mu} S +16\left( \nabla_\mu V^\mu \right)^2\ ,
\label{R2}
\eea
where~$ S^* \overleftrightarrow{\partial_\mu} S =  S^* \partial_\mu S -  S \partial_\mu S^*$.

To construct the supersymmetric $R_{\mu\nu}^{2}$ invariant, we employ the action formula \eq{Phi}. Substituting for the components of the multiplet $(\Phi,\Psi,P)$ given in \eq{com3}, and imposing gauge-fixing conditions \eq{gf}, give the supersymmetric completion of the Ricci tensor squared  as follows
\bea
e^{-1}{\mathcal L}_{ {R_{\mu\nu}^2 + R^2} }  &=& R_{\mu\nu}R^{\mu\nu} -\ft{23}{64}R^{2}-\ft{1}{32} R\left|S\right|^2
-R_{\mu\nu}V^\mu V^\nu+\ft{5}{16} RV^2+\ft{1}{16} (V^2)^2 
\\
&&-\ft{25}{16} V^2 \left|S\right|^{2}  -\ft14 \partial_{\mu}\,S\partial^{\mu}S^*
-\ft{5}{8} \rmi V^\mu S^* \overleftrightarrow{\partial_\mu}S+\ft14 \left(\nabla_\mu V^\mu \right)^{2}
 - F_{\mu\nu} F^{\mu\nu}\ ,
 \nn
\eea
where we have exhibited the bosonic part of the Lagrangian. The $R^2$ dependent part can be removed by adding $\frac{23}{64} {\mathcal L}_{R^2}$ to this Lagrangian, obtaining

\bea
e^{-1}{\mathcal L}_{R_{\mu\nu}^2}  &=& R_{\mu\nu}R^{\mu\nu} -R_{\mu\nu}V^\mu V^\nu + \ft74 RV^2 +\ft{17}{8} R\left|S\right|^2
+\ft{23}{4}|S|^4 - F_{\mu\nu} F^{\mu\nu} \nn
\\
&& +6 \left(\nabla_\mu V^\mu \right)^{2}
+\ft32 (V^2)^2 +\ft{11}{4} V^2 \left|S\right|^{2}  -6\partial_{\mu}S\partial^{\mu}S^*
-\ft{7}{2}\rmi V^\mu S^* \overleftrightarrow{\partial_\mu}S\ .
\eea

Next, we construct the supersymmetric completion of the $RS^n$ term.  To this end, we employ the action formula \eq{Ln}, in which we substitute for the components of the multiplets $(\sigma,\psi,N)$ and $(Z',\Omega',F')$ given in \eq{com4} and \eq{com2}, respectively. Imposing the gauge fixing conditions \eq{gf} in the resulting Lagrangian, and dividing by an overall constant factor of $-(n+1)$, we obtain
\be
e^{-1} \mathcal{L}^{(n)} = \ft12 \left[ R+ \frac{2(3n-1)}{n+1} |S|^2
+2  V^2  -4i  \nabla_{\mu} V^{\mu}\right] S^n  + h.c.\ ,
\label{n}
\ee
where we have given the bosonic part of the result. Note that the $n=0$ case agrees with the Poincar\'e supergravity action \eq{EH} which we obtained by an alternative procedure.

%%%%%%%%%%%%%%%%%%%%%%%%%%%%%%%%%%%%%%
\subsection{$\cN=(1,1)$ Generalized Massive Supergravity}
%%%%%%%%%%%%%%%%%%%%%%%%%%%%%%%%%%%%%%

We now consider  a combination of the invariants up to dimension four, namely,
\be
I = \frac{1}{\kappa^2} \int d^3 x \left[ \ft12 M {\mathcal L}_C +  \sigma {\mathcal L}_{EH}
+ \frac{1}{\mu}  {\mathcal L}_{CS}  + \frac{1}{\nu}  {\mathcal L}_{RS} +\frac{1}{m^2}  {\mathcal L}_{R_{\mu\nu}^{2}} 
+ c_1 {\mathcal L}_{R^2} + c_2{\mathcal L}_{RS^2} \right] \ ,
\label{t1}
\ee
where $(\sigma,M,\mu,\nu, m^2, c_1,c_2)$  are arbitrary real constants. Defining
\be
S= A+\rmi B\ ,
\ee
where $A$ and $B$ are real scalar fields, the $\cN=(1,0)$ supersymmetric truncation is achieved by setting $V_\mu=0$ and $B=0$. In that case the so-called generalized massive gravity (GMG) model is defined by setting
\be
\nu=\infty\ ,\qquad c_1=-\frac{3}{8m^2}\ ,\qquad c_2 = \frac{1}{8m^2}\ .
\label{cc}
\ee
With these choices of the coupling constants the model expanded around supersymmetric $AdS_3$ vacuum propagates only helicity $\pm\, 2$ and $\pm\, 3/2$ states with AdS energies that respect perturbative unitarity. We shall define the $\cN=(1,1)$ supersymmetric version of the GMG model by choosing the coupling constants as in \eq{cc} as well, since the quadratic action obtained by expanding around the supersymmetric $AdS_3$ vacuum contains the $\cN=(1,0)$ sector as an independent subsector. In this case the total Lagrangian becomes
\bea
e^{-1} {\mathcal L}_{GMG} &=&  \sigma (R+2V^2 -2|S|^2 ) + M A
\nn\\
&&-\frac{1}{4\mu} \left[ \epsilon^{\mu\nu\rho}\left( R_{\mu\nu}{}^{ab}\, \omega_{\rho ab}
+ \ft23 \omega_\mu{}^{ab}\, \omega_{\nu b}{}^c\,\omega_{\rho ca} \right)-8\epsilon^{\mu\nu\rho} V_\mu \partial_\nu V_\rho \right]
\nn\\
&& +\frac{1}{m^2} \Big[ R_{\mu\nu} R^{\mu\nu} -\ft38 R^2   - R_{\mu\nu} V^\mu V^\nu -  F_{\mu\nu} F^{\mu\nu} +\ft14 R(V^2 -B^2)
\nn\\
&& \qquad \quad+ \ft16 |S|^2(A^2-4B^2) -\ft12 V^2 (3A^2+4B^2) - 2V^\mu  B\partial_\mu A \Big]\,.
\label{GMG11}
\eea
Remarkably, all terms proportional to $|\partial S|^2, RA^2, (\nabla_\mu V^\mu)^2$ and $(V_\mu V^\mu)^2$ have cancelled. The cancellation of the
$|\partial S|^2$ and $RA^2$ require  $c_1$ and $c_2$ to have the  values given in \eq{cc}, and  it is crucial for having ghost-free propagation of massive modes, as we shall see below.

Notwithstanding that the fields $A$ and $B$ do not propagate, their elimination yields highly nonlinear interactions, including those which take the form of an infinite power series in the Ricci curvature scalar $R$.  In that sense, the notion of a supersymmetric GMG model is extended here, compared to the case of $\cN=(1,0)$ supersymmetric version where the single auxiliary field, a real scalar, can be eliminated from the action by means of its algebraic equation of motion, yielding the standard bosonic GMG action.  Nonetheless, in both cases the action contains the combination $( R_{\mu\nu}R^{\mu\nu} -\frac38 R^2)$, and if we take this feature to be the defining one for an extended definition of super GMG models, it is clear that  such an extension is not unique. In such models, there is no need for eliminating the auxiliary fields, even when they are non-propagating, unless their field equations are algebraic ones.

Turning  to the model with parameters chosen as in \eq{cc}, here we shall focus on maximally supersymmetric AdS vacuum and determine the spectrum of fluctuations around it.  In view of the results of \cite{Deger:2013yla}, the following background is maximally supersymmetric
\be
{\bar R}_{\mu\nu} = -\frac{2}{\ell^2} {\bar g}_{\mu\nu}\ ,\qquad {\bar A} =-\frac{1}{\ell}\  ,\qquad {\bar V}_\mu=0\ ,\qquad \bar B=0\ ,
\label{VEV11}
\ee
where ${\bar g}_{\mu\nu}$ is the $AdS_3$ metric, and $\ell$ is the $AdS_3$ radius which must obey the equation
\be
4\sigma + \ell M + \frac{2}{3\ell^2 m^2} = 0\ .
\ee
Let us define the fluctuation fields around this vacuum as
\bea
g_{\mu\nu}&=& \bar g_{\mu\nu} \left(1+ \ft13 h\right) +  H_{\mu\nu}\ , \qquad \bar g^{\mu\nu} H_{\mu\nu} = 0\ ,
\nn\\
A&=&\bar A + a\ ,\qquad B=\bar B +b\ ,\qquad V_\mu = {\bar V}_\mu + v_\mu\ ,\qquad
\eea
and choose the gauge condition
\be
{\bar\nabla}^\mu H_{\mu\nu}=0\ .
\ee
The linearized field equations then take the form
\bea
&& \left[\cD(1)\,\cD(-1)\,\cD(\eta_+)\, \cD(\eta_-)\, H\right]_{\mu\nu} =  -\frac{1}{3\ell^2} \left(\bar \nabla_\mu \bar \nabla_\nu
 - \frac{1}{3} \bar g_{\mu\nu} \bar \Box \right) h\ ,
\label{om1} \nn\w2
&& \frac{\Omega}{m^2}  \left(\ell^2 {\bar\Box} -3\right)h =0\ ,\qquad   \frac{\Omega}{m^2}\, a=0\ ,
\qquad   \frac{\Omega}{m^2}\, b=0\ ,
\label{om2}\nn\w2
&&  \frac{\Omega}{m^2} \left[ \cD(\eta_+) \,\cD(\eta_-) v\right]_\mu = 0\ ,
\label{om3}
\eea
where
\bea
\eta_\pm = \Omega^{-1} \left(-\frac{\ell m^2}{2\mu}\pm \sqrt{\frac{\ell^2 m^4}{4\mu^2} -\Omega}\right)\ ,\qquad
\Omega \equiv  \sigma\, \ell^2 m^2 - \ft12  \ .
\eea
and $\cD(\eta)$ is a first-order linear differential operator, parametrized by
a dimensionless constant $\eta$, that acts on a rank-$s \ge 1$ totally symmetric, traceless and divergence-free tensor as
\be
\left[ {\cD}(\eta) \,\varphi^{(s)}\right]_{\mu_1\cdots \mu_s}
= \left[{\cD}(\eta)\right]_{\mu_1}{}^\rho \,\varphi^{(s)}_{\rho\mu_2 \cdots\mu_s}\ ,
\qquad
\left[\mathcal{D}\left(\eta\right)\right]_\mu{}^\nu =
\ell^{-1}\,\delta_\mu^\nu + \frac{\eta}{\sqrt{|\bar g|}}\,\varepsilon_\mu{}^{\tau\nu}\bar \nabla_\tau\ .
\ee
The equations for $H_{\mu\nu}$ and $h$  agree precisely with those arising in the $\cN=(1,0)$ GMG model \cite{Bergshoeff:2010mf, Andringa:2009yc} whose spectrum was studied in detail in \cite{Bergshoeff:2010iy}, extending earlier results of \cite{Liu:2009pha} for the bosonic model. For ``non-critical" values of the couplings summarized by the condition $m^{-2} \Omega (\eta_+ - \eta_- ) (|\eta_+|  -1) (|\eta_-| - 1) \ne 0$, these equations describe the UIRs of $SO(2,2)$ with lowest weight $(E_0,s)$, and where $\ell^{-1}E_0$ is the lowest energy, and $s$ is the helicity, their values given by
\be
(E_0,s):\quad (2, 2)\ ,\quad (2,-2)\ ,\qquad \left(1+\frac{1}{|\eta_+|}, \frac{2 \eta_+}{|\eta_+|}\right) \ ,\quad
\left(1+\frac{1}{|\eta_-|},  \frac{2 \eta_-}{|\eta_-|}\right)\ .
\label{spin2}
\ee
The new degrees of freedom arising here furnished by the field $v_\mu$. From \eq{om3} it follows the propagating modes have the representation content
\be
(E_0,s):\quad  \left(1+\frac{1}{|\eta_+|}, \frac{ \eta_+}{|\eta_+|}\right) \ ,\quad
\left(1+\frac{1}{|\eta_-|},  \frac{ \eta_-}{|\eta_-|}\right)\ .
\label{spin1}
\ee
Together with the spin-2 modes displayed in \eq{spin2}, these form the bosonic content of a massive spin-2 supermultiplet of $\cN=(2,0)$ supersymmetry in three dimensions. The structure of this multiplet is similar to the one studied in detail in \cite{Lu:2011mw}. The critical
versions of our $\cN=(2,0)$ GMG model arises for
\be
m^{-2} \Omega (\eta_+ - \eta_- ) (|\eta_+|  -1) (|\eta_-| - 1) = 0\ .
\ee
We shall not examine these points here but we note that the spin-2 sector at critical points has been analyzed in considerable detail in \cite{Bergshoeff:2010mf}.  As for the spin-1 sector, it follows a pattern similar to the one discussed in great detail in \cite{Lu:2011mw}, in the context of a parent supergravity theory whose off-shell degrees of freedom coincide with those of $\cN=(2,0)$ supergravity in three dimensions upon a circle reduction.

%%%%%%%%%%%%%%%%%%%%%%%%%%%%%%%%%%%
\section{${\cN}=(2,0)$  Supergravity Models} \label{section: 4}
%%%%%%%%%%%%%%%%%%%%%%%%%%%%%%%%%%%

This section is devoted to the construction of $\cN=(2,0)$ supergravity invariants. The Poincar\'e supergravity and its cosmological extension has already been given in \cite{Kuzenko:2013uya, Kuzenko:2011rd}, and it is also referred to as Type II minimal supergravity. In this section, we first introduce our gauge fixing choices, and construct the Poincar\'e supergravity and the supersymmetric cosmological constant based on the conformal vector multiplet actions discussed in section \ref{section: 2}. We then proceed to the four-derivative invariants and construct the supersymmetric $R^2$ invariant by the same method. Finally, establishing an analogy between the non-abelian vector multiplet and the Poincar\'e multiplet, we construct the $R_{\m\n}^2$ invariant, and discuss the ghost-free maximally supersymmetric vacuum of the four-derivative extended theory.

%%%%%%%%%%%%%%%%%%%%%%%%%%%%%%%
\subsection{$\cN=(2,0)$ Cosmological Poincar\'e Supergravity}
%%%%%%%%%%%%%%%%%%%%%%%%%%%%%%%

The off-shell Poincar\'e supergravity is obtained from the action formula~\eq{eq:ConfVec} and gauge fixing the superconformal transformations by imposing the following gauge conditions
\be
\rho=1, \qquad \lambda=0, \qquad b_{\mu}=0\ ,
\label{eq:NMGF}
\ee
where the first choice fixes dilatations, the second fixes the $S$-supersymmetry and the third fixes the special conformal symmetry. These gauge choices are maintained provided that
\bea
\Lambda_{D} & = & 0\ ,
\nn\\
\Lambda_{K\mu} & = & -\frac{1}{4}\bar{\epsilon}\,\phi_{\mu}+\frac{1}{4}\bar{\eta}\,\psi_{\mu}+h.c. \ ,
\nn\\
\eta & = & \frac{1}{2} \rmi \gamma\cdot\widehat{G}\,\epsilon+D\epsilon\ .
\eea
We therefore end up with the new minimal Poincar\'e multiplet consisting of a dreibein $e_{\mu}^{a}$, a gravitino $\psi_{\mu}$,
a $U(1)_{R}$ symmetry gauge field $V_{\mu}$, a vector gauge field $C_{\mu}$ and an auxiliary scalar $D$.
%
%\be
%\textrm{New Minimal Poincar\'e multiplet: }\qquad (e_{\mu}^{a},\psi_{\mu},V_{\mu},C_{\mu},D)
%\ee
%
%
The resulting local supersymmetry transformation rules are
\bea \label{nmt}
\delta e_{\mu}{}^{a}& = & \ft12 \,\bar{\epsilon}\,\gamma^{a}\,\psi_{\mu}+h.c. \nn
 \\
\delta\psi_{\mu}& = & \left(\partial_{\mu}+\ft{1}{4}{\omega}_{\mu}{}^{ab}\,\gamma_{ab}
-\rmi V_{\mu}\right)\epsilon-\ft{1}{2} \rmi\, \gamma_{\mu}\gamma\cdot\widehat{G}\epsilon-\gamma_{\mu}\,D\epsilon \nn
\\
\delta C_{\mu}& = & -\ft{1}{4}\rmi \,\bar{\epsilon}\,\psi_{\mu}+h.c. \nn
\\
\delta V_{\mu}& = & -\ft{1}{2}\rmi \bar{\epsilon}\,\gamma^{\nu}\widehat{\psi}_{\mu\nu} +\ft{1}{8} \rmi \bar{\epsilon}\,\gamma_{\mu}\gamma\cdot\widehat{\psi}
-\ft{1}{2} \bar{\epsilon}\, \gamma\cdot\widehat{G}\,\psi_\mu +\rmi D\bar\epsilon\, \psi_\mu +h.c. \nn
 \\
\delta D & = & -\ft{1}{16}\bar{\epsilon}\,\gamma\cdot\widehat{\psi}+h.c.
\eea
where  the $U(1)_{R}$ covariant gravitino field strength is given by
\be
\widehat{\psi}_{\mu\nu}=2\left(\partial_{[\mu}+\ft{1}{4}{\omega}_{[\mu|}{}^{ab}\,\gamma_{ab}
-\rmi V_{[\mu}\right)\psi_{\nu]}-\rmi  \gamma_{[\mu}\gamma\cdot\widehat{G}\psi_{\nu]}-2D\,\gamma_{[\mu}\psi_{\nu]}\ .
\ee
Substituting the gauge fixing conditions (\ref{eq:NMGF}) into the Lagrangian (\ref{eq:ConfVec}), and rescaling with a factor of $-16$,
we obtain the following Poincar\'e supergravity
\be
e^{-1}\mathcal{L}_{EH}=R -2 G^2-8D^{2}-8\epsilon^{\mu\nu\rho}\,C_{\mu}\,\partial_{\nu}V_{\rho}\ ,
\label{eq:NMPoincare}
\ee
where we have defined
 \be
 G_\mu := \e_{\mu\nu\rho} G^{\nu\rho}\ , \qquad    G^2 := G_\mu G^\mu\,.
 \ee
Consequently, $G_\m$ is a covariantly conserved tensor $\nabla^\m G_\m = 0$. A supersymmetric cosmological constant can be added to the Poincar\'e supergravity (\ref{eq:NMPoincare}), which can be obtained from the action formula \eq{Dprime}, and imposing the gauge fixing choices~(\ref{eq:NMGF}), obtaining
\be
e^{-1}\mathcal{L}_{C} = 2D-\epsilon^{\mu\nu\rho}\,C_{\mu}\,G_{\nu\rho} - \left(\frac18\bar{\psi}_{\mu}\,\gamma^{\mu\nu}\,\psi_{\nu}+h.c.\right)\ .
\ee

%%%%%%%%%%%%%%%%%%%%%%%%%%%%%%%%%%%%%%%%%%%%%%
\subsection{$\cN=(2,0)$  $RD$ and $R^2$ Invariants}
%%%%%%%%%%%%%%%%%%%%%%%%%%%%%%%%%%%%%%%%%%%%%%
For the construction of the $RD$ invariant, we consider the vector multiplet action (\ref{Dprime}) for the primed vector multiplet $(\r^\prime, C_\m^\prime, \l^\prime, D^\prime )$. Using the composite expressions given in \eq{nc1} and fixing the redundant superconformal symmetries by using the gauge fixing choices \eq{eq:NMGF}, give the supersymmetric completion of the $RD$ action
\be
e^{-1} \cL_{RD} = RD + 8 D^3  - 2 G^{\mu\nu}\left( F_{\m\n} + \nabla_\mu G_\nu  +2 D G_{\mu\nu}\right) + \ft12 \epsilon^{\m\n\r}\, V_\m\, F_{\n\r} \ ,
\label{RD}
\ee
where we have rescaled the Lagrangian with an overall factor of $-8$. Note that although the $RD$ invariant and the Lorentz-Chern-Simons invariant (\ref{CS1}) have the same conformal $\epsilon^{\m\n\r} V_{\m} F_{\n\r}$ term, the $RD$ invariant is not conformally invariant as can be understood from the existence of the Ricci scalar. Such non-conformal invariants are studied in detail in the context of Chern-Simons contact terms in three dimensions \cite{Closset:2012vp,Closset:2012vg,Closset:2012ru}.

Next, we construct the supersymmetric completion of $R^2$. Using the composition formula \eq{nc1}  and employing the gauge fixing choices (\ref{eq:NMGF}) in the action formula \eq{PreYM},  we obtain
\be
e^{-1}\mathcal{L}_{R^{2}}  =  (R + 24 D^2 + 2G^2 )^2  -8 \left( F_{\mu\nu} + 2\nabla_{[\mu} G_{\nu]} +4 DG_{\mu\nu} \right)^2
+  64D\Box D\ .
\label{R220}
\ee
%

%%%%%%%%%%%%%%%%%%%%%%%%%%%%%%%%%%%%%%%%%%%
\subsection{$\cN=(2,0)$ $R_{\mu\nu}^2$ Invariant} \label{ss: 43}
%%%%%%%%%%%%%%%%%%%%%%%%%%%%%%%%%%%%%%%%%%%

The supersymmetric completion of the Ricci tensor-squared term is most conveniently obtained by establishing a map between Yang-Mills and supergravity multiplets. To do so, we begin by gauge fixing the nonabelian version of the transformation rules \eq{vtr} in accordance with  (\ref{eq:NMGF}), obtaining
\bea
 \delta C_{\mu}^I &=& \ft{1}{2}\bar{\epsilon}\, \gamma_{\mu} \lambda^I - \ft{1}{4} \rmi \rho^I \bar{\epsilon}\, \psi_{\mu} + h.c. \ ,
 \nn\\
 \delta \rho^I &=&  \rmi \bar{\epsilon} {\l}^I + h.c.  \ ,
 \nn\\
 \delta \lambda^I &=&  - \ft{1}{4} \gamma^{\mu\nu} \widehat{G}^I_{\mu\nu} \epsilon +  \ft{1}{2} \rmi D^I \epsilon
 - \ft{1}{4}  \rmi \slashed{\widehat{D}} \rho^I \epsilon -  \ft{1}{2} \rmi \rho^I D \e + \ft14 \r^I \g \cdot \widehat{G} \e    \ ,
 \nn\\
 \delta D^I &=& ( - \ft{1}{2} \rmi \bar{\epsilon}\, \slashed{\widehat{D}} \lambda^I+ \ft{1}{2} \rmi D \bar\e \lambda^I - \ft14 \bar\e\, \g \cdot \widehat{G} \l^I + \ft14 g \, \bar\e\, f_{JK}{}^I\, \r^J\l^K + h.c. )  \,.
\eea
where
\bea
 \widehat{D}_{\mu}\rho^I & = & \partial_{\mu}\rho^I + \left(-\rmi \bar{\psi}_{\mu}\,\lambda^I + h.c. \right) + g\, f_{JK}{}^I\, C_\m^J\,\r^K\ ,
 \nn\\
 \widehat{D}_{\mu}\lambda^I & = & (\partial_{\mu} + \ft14 \o_\m{}^{ab}\, \g_{ab} - \rmi V_{\mu}) \lambda^I
 + \ft{1}{4} \gamma^{\rho\sigma} \widehat{G}^I_{\rho\sigma}\, \psi_{\mu}  - \ft{1}{2} \rmi D^I \psi_{\mu}  + \ft{1}{4} \rmi \slashed{\widehat{D}} \rho^I \psi_{\mu}   \nn\\
&&
+  \ft{1}{2} \rmi \rho^I D\, \psi_\m - \ft14 \r^I \g \cdot \widehat{G} \psi_\m + g\, f_{JK}{}^I C_\m^J\, \l^K\ , \nn\\
{\widehat{G}_{\mu\nu}^I } & = & 2 \partial_{[\mu} C_{\nu]}^I - \left( \bar{\psi}_{[\mu} \gamma_{\nu]} \lambda^I - \ft{i}{2} \rho^I \bar{\psi}_{\mu}\, \psi_{\nu} + h.c. \right) + g \, f_{JK}{}^I\, C_\m^J\, C_{\nu}^K \, .
\label{PYM}
\eea
We will next show that the following set of fields
\bea
 ( \O_\m{}^{- ab}\,, \widehat{G}^{ab}\,,  \widehat{\p}^{ab}\,, \widehat{F}^{ab}(V_+,\o, \O^-) )
\eea
transform as a Yang-Mills multiplet $ (C_\m^I\,, \rho^I \,, \l^I \,, D^I)$, where the $ab$ index pair plays the role of Yang-Mills index. The definitions of the torsionful spin connection $\O_\m{}^{- ab}$, the gravitino field strength $\widehat\p_{ab}$, and the modified $U(1)_R$ gauge field are given by
\bea
\O_\m{}^{ab\, \pm} &=& \o_\m{}^{ab} \pm 2 \ve_\m{}^{ab} D \ ,
\label{nd1}\\
\widehat{\p}_{ab} &=& 2 \nabla_{[a} (\o, \O^+ , V ) \p_{b]} - \rmi \g_{[a} \g \cdot \widehat{G} \p_{b]} \ ,
\label{nd2}\\
V_{a +} &=& V_a + \ft12 \e_{a}{}^{bc} \widehat{G}_{bc} \ ,
\label{nd3}
\eea
where in the definition of $\widehat\p^{ab}$, the connection $\o$ rotates the Lorentz vector index while the connection $\O^+$ rotates the Lorentz spinor index.

First, we calculate the transformation rules for ${\o}_\m{}^{ab}$, $D$ and $\widehat{G}_{ab}$
\bea
\d \o_\m{}^{ab} &=& -\ft14 \bar\e\, \g_\m \widehat\p_{ab} + \ft12 \bar\e\, \g^{[a} \widehat\p^{b]}{}_\m +  D \bar\e\, \g_{ab}\, \p_\m - \rmi \bar\e\, \p_\m \widehat{G}^{ab} + h.c. \ ,
\label{mt1}\\
\d D &=&-\ft{1}{16}\bar{\epsilon}\,\gamma\cdot\widehat{\psi}+h.c. \ ,
\label{mt2}\\
\d \widehat{G}_{ab} &=& - \ft14 \rmi \bar\e\, \widehat{\p}_{ab} + h.c. \ .
\label{mt3}
\eea
From the first two equations, we observe that
\bea
\d \O_\m{}^{-ab} &=& - \ft12 \bar\e\, \g_\m \widehat{\p}^{ab} -  \rmi \bar\e\, \p_\m \widehat{G}^{ab} + h.c.  \ .
\label{otr}
\eea
Next, we compute the transformation rule for the gravitino curvature
\bea
\d \widehat\p_{ab} &=& \ft14 \g^{cd}\, \widehat{R}_{abcd}(\O^+) \e  -  \rmi \widehat{F}_{ab}(V) \e - 2 \rmi \nabla_{[a} (\o ) \widehat{G}_{b] c} \g^c \e \nn\\
&& - \rmi \nabla_{[a} (\o)\, \widehat{G}^{cd} \ve_{b]cd} + 2 \rmi D\, \widehat{G}_{ab} \e  - \widehat{G}_{ab} \g \cdot \widehat{G} \e \nn\\
&& + \rmi \widehat{G} \g_{ab}\, \g \cdot \widehat{G} \e \,,
\eea
where $\widehat{R}_{abcd}(\O^+)$ represents a torsionful supercovariant Riemann tensor. Using the definition of $V_+$ given in (\ref{nd3}),  the Bianchi identity $\nabla_{[a} \widehat{G}_{bc]} = 0$ and $\widehat{R}_{abcd}(\O^+) = \widehat{R}_{cdab}(\O^-)$, we rewrite the transformation rule for the gravitino curvature as
\bea
\d \widehat\p_{ab} &=& \ft14 \g^{cd}\, \widehat{R}_{cdab}(\O^-) \e  -  \rmi \widehat{F}_{ab}(V_+) \e + \rmi \slashed{\nabla}(\O^-) \widehat{G}_{ab}\, \e  - \widehat{G}_{ab}\, \g \cdot \widehat{G} \e \ ,
\eea
where in $\nabla_{\m} (\O^-) \widehat{G}_{ab}$, the connection $\O^-$ rotates both $a$ and $b$ indices. Finally, defining
$\widehat{F}_{ab} (V_+, \o, \O^-)$ where  $\o$ rotates the Lorentz vector index $b$, whereas the connection $\O^-$ rotates the index $c$ in the covariant derivative acting on $\widehat{G}_{bc}$, we have
\bea
\d \widehat\p_{ab} &=& \ft14 \g^{cd}\, \widehat{R}_{cdab}(\O^-) \e  -  \rmi \widehat{F}_{ab}(V_+, \o , \O^-) \e + \rmi \slashed{\nabla}(\O^-) \widehat{G}_{ab}\, \e  \nn\\
&&  - \widehat{G}_{ab}\, \g \cdot \widehat{G}\, \e + 2 \rmi D G_{ab}\, \e \,.
\label{gtr}
\eea
Finally, we consider the transformation rule for $\widehat{F}_{ab}(V_+, \o , \O^-)$
\bea
\d \widehat{F}_{ab}(V_+, \o , \O^-)  = \ft14 \rmi \bar\e\, \slashed{\nabla}(\o, \O^-) \widehat\p_{ab} - \ft14 \rmi D \bar\e \,\widehat\p_{ab}  + \ft1{8} \bar\e\, \g \cdot \widehat{G}\, \widehat{\p}_{ab}  -  \rmi \bar\e\, \widehat{G}_{c[a} \widehat\p_{b]}{}^c+ h.c.  \,,\quad
\label{ftr}
\eea
where in $\nabla_c (\o, \O^-) \widehat\p_{ab}$ the connection $\o$  acts on the spinor index, whereas $\O^-$ acts on both $a$ and $b$ indices.

Comparing the transformation rules \eq{mt3}, \eq{otr}, \eq{gtr} and \eq{ftr} with those of the nonabelian vector multiplet, we find the following correspondence
\bea
\O_\m{}^{-ab} \leftrightarrow C_\m^I\,,\quad  4 \widehat{G}^{ab} \leftrightarrow \r^I\,,\quad
- \widehat\p^{ab} \leftrightarrow \l^I \,, \quad 2 \widehat{F}^{ab}(V_+,\o, \O^-)  \leftrightarrow  D^I  \,.
\label{YMMap}
\eea

We now turn to the supersymmetric completion of the Ricci squared term. To this end, we first construct the following Lagrangian
\bea
e^{-1} \cL_{YM} &=&  \ft14 \Big( G_{\m\n}^I - \r^I G_{\m\n} \Big) \Big( G^{\m\n I} - \rho^I G^{\m\n} \Big) \nn\\
&& - \ft12 (D^I - \rho^I D)^2 + \ft18 D_\m \r^I D^\m \r^I \,,
\eea
describing the bosonic sector of Yang-Mills multiplet coupling to supergravity. This is obtained by generalizing the superconformal invariant action (\ref{PreYM})  and then fixing gauges according to (\ref{eq:NMGF}).
It is now straightforward to use the map (\ref{YMMap}) which gives the bosonic part of the supersymmetric completion of the Riemann squared action
\bea
e^{-1} \cL_{Riem^2} &=&  \ft14 \Big( R_{\m\n ab} (\Omega^-)  - 4G_{a b}\, G_{\m\n} \Big) \Big( R^{\m\n ab} (\O^-) - 4 G^{ab}\, G^{\m\n} \Big)
\nn\\
&& - 2 \Big( F_{ab}(V_+,\o, \O^-) - 2 D G_{ab} \Big) \Big(F^{ab}(V_+,\o,\O^-) - 2 D G^{ab} \Big) \nn\\
&& + 2 \nabla_\m (\O^-) G_{ab}\, \nabla^\m (\O^-) G^{ab} \,.
\eea
Finally, expanding the torsion terms and using the definition of three-dimensional Riemann tensor
\bea
R_{\m\n ab} = \ve_{\m\n\r}\, \ve_{abc} \Big( R^{\r c} - \ft12 e^{\r c} R \Big) \,,
\eea
we obtain the supersymmetric completion of the Ricci squared action
\bea
e^{-1} \cL_{R_{\m\n}^2} &=& R_{\m\n}\, R^{\m\n} - \ft14 R^2 + 4 R D^2   +RG^2 -2R_{\mu\nu}\,G^\mu\, G^\nu + 48 D^4 + 8 D \Box D
\nn\\
&& +8D^2G^2 +(G^2)^2  - 2 ( F_{\mu\nu}  + \nabla_{[\mu} G_{\nu]}  )^2- \left(\nabla_\mu G_\nu + 4DG_{\mu\nu} \right)^2\ ,
\label{Ric2}
\eea
where we recall that  $G_\mu := \e_{\mu\nu\rho}\, G^{\nu\rho}$.  If desired, a term proportional to ${\cal L}_{R^2}$ from \eq{R220} can be added to this result to obtain the invariant in which  the only curvature squared term is that of the Ricci tensor.

We conclude this subsection with comments on the existence of an off-shell $RD^2$ invariant. Considering the  vector multiplet action (\ref{eq:VecGen}) and the composite formulae~(\ref{nc1}) and~(\ref{nc2}), we find the following choices for $C_{IJ}$ to obtain a supersymmetric completion for the $RD^2$ term:
\begin{enumerate}

\item{The supersymmetric completion of the $RD^2$ term can be obtained by supersymmetrizing the $\Box^c (\r^{-3} D^2)$ term. In order to do so, we can consider two vector multiplets:~$(\r, C_\m, \l, D)$ labeled by~$1$, and~$(\r^{\prime\prime}, C^{\prime\prime}_\m, \l^{\prime\prime}, D^{\prime\prime})$ labeled by~$3$, and set~$C_{13} = \r^{-1}$. Making this choice, we find that all the terms in the Lagrangian (\ref{eq:VecGen}) cancel each other out, thus, not giving rise to an $RD^2$ invariant.}

\item{Alternatively, one can consider the supersymmetric completion of $\r^{-2} D^2 \Box^c \r$ which gives rise to an $RD^2$ term after gauge fixing. Such a model can be obtained by considering two vector multiplets: $(\r, C_\m, \l, D)$ labeled by~$1$, and $(\r^{\prime}, C^{\prime}_\m, \l^{\prime}, D^{\prime})$ labeled by~$2$, and set $C_{22} = \r^{-1}$. Making this choice, however, we find that the resulting action is the $R^2$ action given in (\ref{R220}).}

\item{Another alternative is the supersymmetric completion of $\r^{-2} D \Box^c (\r^{-1} D)$. This construction also corresponds to the choice $C_{22} = \r^{-1}$, and coincides with the $R^2$ action given in (\ref{R220}) }
\end{enumerate}
In view of these arguments, it is not clear to us how the supersymmetric completion of $RD^2$ as an off-shell invariant independent of the $R^2$ and $R_{\mu\nu}^2$ invariants can be obtained within the tensor calculus framework presented in section \ref{section: 2}.

%%%%%%%%%%%%%%%%%%%%%%%%%%%%%%%%%%%%%%
\subsection{$\cN=(2,0)$ Generalized Massive Supergravity}
%%%%%%%%%%%%%%%%%%%%%%%%%%%%%%%%%%%%%%

We now consider  a combination of the invariants unto dimension four, namely,
\be
I = \frac{1}{\kappa^2} \int d^3 x\,  \left[ M {\mathcal L}_C +  \sigma {\mathcal L}_{EH}
+ \frac{1}{\mu}  {\mathcal L}_{CS}   + \frac1{\nu} \cL_{RD}  +\frac{1}{m^2}  {\mathcal L}_{R_{\mu\nu}^{2}} + c \, {\mathcal L}_{R^2} \right] \ ,
\label{t2}
\ee
where $(\sigma,M,\mu, \n, m^2, c)$  are arbitrary real constants. This action is invariant under the off-shell supersymmetry transformation rules given in \eq{nmt}. If we consider the defining feature of a super GMG model to be that it contains the term $R_{\mu\nu}R^{\mu\nu} -\frac38 R^2$, such an extension is clearly not not unique, as discussed earlier. Focusing on maximally supersymmetric backgrounds and ghost free fluctuations around it, we begin by noting that the metric for such backgrounds is AdS or Minkowski. In the former case, $D$ must be non-vanishing, and this is problematic for ghost-freedom due the presence of the $RD^2$ term in the action. Such a term is akin to the $RA^2$ term in the $\cN=(1,1)$ model which we are able to eliminate. In the case of Minkowski background, the presence of the $RD^2$ term is harmless. Thus, to achieve maximally supersymmetric Minkowski
background, we are led to consider the model with the following choice of parameters
\be
M=0\ ,\qquad \n = \infty\ , \qquad c =-\frac{1 }{8m^2}\ .
\label{cc2}
\ee
In this case,  the total Lagrangian becomes
\bea
 e^{-1} {\mathcal L}_{GMG} &=& \sigma \, (R - 2G_\mu G^\mu  - 8D^2  - 4G^\mu V_\mu )
\nn\\
&&-\frac{1}{4\mu}\epsilon^{\mu\nu\rho} \left[  R_{\mu\nu}{}^{ab}\, \omega_{\rho ab}
+ \ft23 \omega_\mu{}^{ab}\, \omega_{\nu b}{}^c\,\omega_{\rho ca} - 8V_\mu \partial_\nu V_\rho
\right] \nn
\\
&& + \frac{1}{m^2} \Big[ R_{\mu\nu} R^{\mu \nu} - \frac{3}{8} R^2
 - 2 R D^2 -  R_{\mu\nu}\, G^\mu G^\nu  +\frac12 RG^2
\\
 && - 24 D^4    +\frac12 (G^2)^2    - 4 D^2 G^2  - F_{\mu\nu} F^{\mu\nu} \Big.
 + 8 D G^{\mu\nu} \left( F_{\mu\nu}  + \nabla_\mu G_\nu\right) \Big] . \nn
\eea
For the maximally supersymmetric background, the fields $(D,V_\mu, C_\mu)$ are vanishing. Therefore, the analysis of the linearized fluctuations for spin-2 modes around this  background is the same as that of standard GMG model, amounting to the purely gravitational part of the action above. Thus, we know that the system describes two massive helicity $\pm 2$ modes with masses~\cite{Ohta:2011rv}
\be
m^2_\pm = - \sigma m^2 +\frac{m^4}{2\mu^2} \left[ 1\pm \sqrt{1-  \frac{4\sigma\mu^2}{m^2}}\right]\ .
\label{mf2}
\ee
Ghosts are absent for $m^2>0$ and $\sigma \le 0$ \cite{Bergshoeff:2009hq,Andringa:2009yc,Ohta:2011rv}.  Next, we note that the linearized fluctuation of the field $D$ vanishes. Denoting the linearized vector fluctuations of $(V_\mu, C_\mu)$ by the same symbols and choosing the Lorentz gauges $\partial_\mu V^\mu=0$ and $\partial_\mu C^\mu=0$, one finds that their linearized field equations are
\be
\frac{1}{m^2} \Box V^\mu -\frac{1}{\mu} \epsilon^{\mu\nu\rho} \,\partial_\nu V_\rho + \sigma G^\mu =0\ ,
\qquad F_{\mu\nu} +2 \partial_{[\mu} G_{\nu]} = 0\ .
\ee
A simple manipulation of these equations gives
\be
\left[\,(\Box +\sigma m^2) \delta_\mu^\rho\, \delta_\nu^\sigma  -\frac{\sigma m^2}{\mu} \epsilon_{[\mu}{}^{\rho\sigma} \partial_{\nu]}\,\right]
\begin{pmatrix} F_{\rho\sigma}\\ G_{\rho\sigma}\end{pmatrix} =0\ ,
\ee
Diagonalizing the mass matrix one finds that the masses for $V^\mu$ and $C^\mu$ are given by the formula \eq{mf2}.  Thus, we have found the bosonic sector of two massive spin-2 multiplets of $\cN=(2,0)$ supersymmetry.

%%%%%%%%%%%%%%%%%%%%%%%%%%
\section{Conclusions} \label{section: 5}
%%%%%%%%%%%%%%%%%%%%%%%%%%
In this paper we have completed the construction of all off-shell Poincar\'e supergravity invariants up to mass dimension four and with~$\mathcal{N}=(1,1)$ and $\mathcal{N}=(2,0)$ supersymmetry. We have mostly utilized superconformal tensor calculus except for the supersymmetric completion of Ricci tensor squared invariant with~$\mathcal{N}=(2,0)$ supersymmetry, where we have employed a map between the Yang-Mills multiplet and the Poincar\'e multiplet. The resulting Lagrangians with~$\mathcal{N}=(1,1)$ and~$\mathcal{N}=(2,0)$ supersymmetry contain seven and six free parameters respectively, each of which corresponds to separate off-shell invariants. We have determined the relation between the parameters so that the spectrum of fluctuations about a maximally symmetric vacuum solution is ghost-free.  For ghost-free fluctuations about $AdS_{3}$ vacuum, certain type of off-diagonal invariants with mass dimension four, namely $RS^{2}$ for~$\mathcal{N}=(1,1)$ supersymmetry and~$RD^{2}$ for~$\mathcal{N}=(2,0)$ supersymmetry, without curvature squared terms in their supersymmetric completion, play a crucial role. We have constructed the former, but surprisingly we have found that the latter does not seem to exist. Consequently, the $\cN=(2,0)$ model does not seem to have a supersymmetric AdS vacuum with ghost-free spectrum, even though it does admit a supersymmetric Minkowski vacuum that gives ghost-free massive spin-2 multiplet.

There are a number of directions to pursue in probing the properties of the general class of off-shell supergravities constructed here. While the supersymmetric AdS vacuum solutions were examined here, it will be instructive to study the non-supersymmetric $AdS$ vacuum solutions as well. A systematic study of the ghost-free vacua and their stability under quantum corrections would also be useful. Such studies would also shed light on the role of  extended supersymmetry and the differences between the two versions of the off-shell $\cN=2$ theory at the quantum level.

Although we constructed vector multiplet actions by using an arbitrary function of vector multiplet scalars, as given in (\ref{eq:VecGen}}, we did not consider such constructions for the scalar multiplet in this paper.It would be interesting to consider the coupling of an arbitrary number of scalar multiplets and vector multiplets, since that would enable us to construct a large class of supergravity Lagrangians \cite{deWit:2006gn}. The composite expression we derived for both scalar and  vector multiplet can also be used to construct matter-coupled higher derivative supergravity models. Such three dimensional matter coupled theories have attracted considerable amount of attention in the context of rigid supersymmetric theories on three-manifolds \cite{Festuccia:2011ws,Closset:2012ru}. Since the compensating multiplet used in the construction of~$\mathcal{N}=(1,1)$ theory includes a complex scalar, our gauge choice fixes R-symmetry in addition to dilatations. However, this is not the case for the ~$\mathcal{N}=(2,0)$ theory, since we gauge fix dilatations with a real scalar. Therefore, one can use this setup to obtain an Einstein-Maxwell theory where the R-symmetry is dynamically gauged.

Finally, we would like to mention that since $\cN=(1,1)$ generalized massive gravity admits a maximally supersymmetric $AdS_3$ vacuum, one should expect a holographically dual superconformal field theory. Here, we do not attempt to calculate the central charge as in the original argument of Brown-Henneaux \cite{Brown:1986nw}, but consider the bosonic truncation of the $\cN=(1,1)$ generalized massive gravity (\ref{GMG11}) as in \cite{Bergshoeff:2010mf,Andringa:2009yc}. For parity-preserving theories with higher derivative extensions, the left and right central charges are given by \cite{Saida:1999ec,Kraus:2005vz}
\bea
c_L = c_R = \frac{\ell}{2G_3} g_{\m\n} \frac{\partial (e^{-1} \cL)}{\partial R_{\m\n}} \ .
\eea
Parity-violating terms result in a difference in the left and right central charges. Given the Lagrangian \eq{t1} with parameter choices \eq{cc}, the only parity violating contribution comes from the Lorentz Chern-Simons term and it is given by $\pm \ft{3}{2G_3 \m}$.Therefore, the central charges read
\bea
c_{L,R} = \frac{3\ell}{2G_3} \Big(\s + \frac{1}{2 m^2 {\ell}^2} \pm \frac{1}{\m \ell} \Big) \,.
\eea
Note that this result precisely matches with the three dimensional $\cN = (1,0)$ model \cite{Andringa:2009yc} since vacuum expectation values for the R-symmetry gauge field $V_\m$ and the imaginary part of the auxiliary scalar $S$ vanish (\ref{VEV11}).

%%%%%%%%%%%%%%%%%%%%
\subsection*{Acknowledgments}
%%%%%%%%%%%%%%%%%%%%

E.S. would like to thank Groningen University for hospitality  and Yi Pang for useful discussions. The research of E.S. is supported in part by NSF grant PHY-1214344. L.B. is supported by the Dutch stichting voor Fundamenteel Onderzoek der Materie (FOM). G.A. acknowledges support by a grant of the Dutch Academy of Sciences (KNAW)

\appendix

%%%%%%%%%%%%%%%%%%%%%%%
\section{Complex Spinor Conventions}\label{section: A}
%%%%%%%%%%%%%%%%%%%%%%%

The metric signature is $\left(-,+,+\right)$. The gamma matrices satisfy the Clifford algebra, i.e. ${\left\{ \gamma^{a},\gamma^{b}\right\}=2\eta^{ab}}$, and the identities
\be
\left(\gamma^{\mu}\right)^{\dagger}=\gamma^{0}\gamma^{\mu}\gamma^{0},\qquad\left(\gamma^{\mu}\right)^{T}=-C\gamma^{\mu}C^{-1},\qquad\left(\gamma^{\mu}\right)^{*}=B\gamma^{\mu}B^{-1}\ ,
\ee
where $C$ is the charge conjugation matrix and $B$ is a unitary matrix with properties
\be
CC^{\dagger}=1\ ,   \qquad   CC^{*}=-1\ ,  \qquad   C^{T}=-C\ .
\ee
\be
C=iB\gamma^{0}\ ,  \qquad BB^{\dagger}=1\ ,  \qquad BB^{*}=1,\qquad B^{T}=B\ .
\ee
For Dirac spinors, there are two different definitions of the conjugate which are given by \cite{Cecotti:2010dg}
\be
\bar{\epsilon}=\rmi \epsilon^{\dagger}\gamma^{0},\qquad\tilde{\epsilon}=\overline{\left(B\epsilon\right)^{*}}\ .
\ee
For Majorana spinors, we impose the reality condition $\epsilon^{*}=B\epsilon$ and Majorana conjugation $\bar{\epsilon}=\epsilon^{T}C$ is equivalent
to Dirac conjugation $\bar{\epsilon}=\rmi\epsilon^{\dagger}\gamma^{0}$.

In order to obtain the flipping rules for bilinears formed by Dirac spinors, it is useful to decompose a Dirac spinor into two Majorana
spinors as $\epsilon_{D}=\epsilon_{M1}+ \rmi\epsilon_{M2}$. As a result, we have
\bea
\left(B\epsilon_{D}\right)^{*} & = & \epsilon_{M1}-\rmi\epsilon_{M2}\ ,\qquad\bar{\epsilon}_{D}
=\bar{\epsilon}_{M1}-\rmi\bar{\epsilon}_{M2},\qquad\tilde{\epsilon}_{D}=\bar{\epsilon}_{M1}+\rmi\bar{\epsilon}_{M2}\ ,
\eea
from which one can obtain
\be
\bar{\epsilon}_{1}\Gamma\left(B\epsilon_{2}\right)^{*}=\alpha\,\bar{\epsilon}_{2}\Gamma\left(B\epsilon_{1}\right)^{*},\qquad\tilde{\epsilon}_{1}\Gamma\epsilon_{2}=\alpha\,\tilde{\epsilon}_{2}\Gamma\epsilon_{1}\ ,
\ee
where $\Gamma$ is any element of the Clifford algebra and $\alpha$ is the corresponding numerical factor in the Majorana flipping relations.
Using the decomposition, one also gets
\be
\tilde{\epsilon}_{1}\Gamma\left(B\epsilon_{2}\right)^{*}=\alpha\,\bar{\epsilon}_{2}\Gamma\epsilon_{1}\ .
\label{eq:notflipping}
\ee
Note that this time we get a different type of bilinear, which becomes an important issue in the closure of the algebra on the scalar multiplet.
Namely, $QQ$ commutation leads to a translation parameter
\be
\xi_{3}^{\mu}=\frac{1}{2}\tilde{\epsilon}_{2}\gamma^{\mu}\left(B\epsilon_{1}\right)^{*}
-\frac{1}{2}\tilde{\epsilon}_{1}\gamma^{\mu}\left(B\epsilon_{2}\right)^{*}\ ,
\ee
which can be shown to be identical to the usual translation parameter
\be
\xi_{3}^{\mu}=\frac{1}{2}\bar{\epsilon}_{2}\gamma^{\mu}\epsilon_{1}-\frac{1}{2}\bar{\epsilon}_{1}\gamma^{\mu}\epsilon_{2}\ ,
\ee
by using \eqref{eq:notflipping}.

The charge conjugation of a spinor is defined by $\lambda^{C}=B^{-1}\lambda^{*}=\left(B\lambda\right)^{*}$ and the complex conjugation of bilinears are
\bea
\left(\bar{\chi}\Gamma\lambda\right)^{*} & \equiv & \left(\bar{\chi}\Gamma\lambda\right)^{C}=\overline{\chi^{C}}\Gamma^{C}\lambda^{C}=\tilde{\chi}\Gamma^{C}\left(B\lambda\right)^{*}\ ,
\eea
\be
\left(\tilde{\chi}\Gamma\lambda\right)^{*}\equiv\left(\tilde{\chi}\Gamma\lambda\right)^{C}
=\widetilde{\chi^{C}}\Gamma^{C}\lambda^{C}=\bar{\chi}\Gamma^{C}\left(B\lambda\right)^{*}\ ,
\ee
where the charge conjugation of matrices are determined by $\left(\Gamma_{1}\Gamma_{2}\right)^{C}=\Gamma_{1}^{C}\Gamma_{2}^{C}$
and $\gamma_{\mu}^{C}=\gamma_{\mu}$.

%%%%%%%%%%%%%%%%%%%%%%%%%
\section{Fierz Identities}
%%%%%%%%%%%%%%%%%%%%%%%%%

Elements of the Clifford algebra in $3D$ are $\left\{ \Gamma^{A}=\mathds{1},\gamma^{\mu}\right\} $ with the orthogonality relation $Tr\left(\Gamma^{A}\Gamma_{B}\right)=2\,\delta_{B}^{A}$. Therefore, any $2$-dimensional matrix can be expanded in the basis
$\left\{ \Gamma^{A}\right\} $ as $M=\frac{1}{2}\sum_{A}\textrm{Tr}\left(M\Gamma_{A}\right)\Gamma^{A}$. As a result, the Fierz identity in $3D$ is given by
\be
\bar{\chi}_{1}\,\chi_{2}\,\epsilon=-\frac{1}{2}\left(\bar{\chi}_{1}\,\epsilon\,\chi_{2}+\bar{\chi}_{1}\,\gamma^{a}\epsilon\,\gamma_{a}\chi_{2}\right)\ ,
\label{eq:fierz}
\ee
from which one can also obtain
\bea
\bar{\chi}_{1}\,\gamma^{a}\chi_{2}\,\gamma_{a}\epsilon & = & -\bar{\chi}_{1}\,\chi_{2}\,\epsilon-2\bar{\chi}_{1}\epsilon\,\chi_{2}\ ,
\label{eq:gamma1}\\
\bar{\chi}_{1}\,\gamma^{ab}\chi_{2}\,\gamma_{ab}\epsilon & = & 2\bar{\chi}_{1}\,\chi_{2}\,\epsilon+4\bar{\chi}_{1}\,\epsilon\,\chi_{2}\ .
\label{eq:gamma2}
\eea
Whenever flipping relations are applicable, one can also obtain additional identities by antisymmetrizing (\ref{eq:gamma1})--(\ref{eq:gamma2})
with respect to $1\longleftrightarrow2$
\bea
\tilde{\chi}_{1}\,\gamma^{a}\chi_{2}\,\gamma_{a}\epsilon & = & -\tilde{\chi}_{1}\,\epsilon\,\chi_{2}+\tilde{\chi}_{2}\,\epsilon\,\chi_{1}\ ,
\label{eq:gamma1flipped}\\
\tilde{\chi}_{1}\,\gamma^{ab}\chi_{2}\,\gamma_{ab}\epsilon & = & 2\tilde{\chi}_{1}\,\epsilon\,\chi_{2}-2\tilde{\chi}_{2}\,\epsilon\,\chi_{1}\ ,
\eea
which are also true for bilinears of type $\bar{\epsilon}_{1}\Gamma\left(B\epsilon_{2}\right)^{*}$. Using \eqref{eq:gamma1flipped} in \eqref{eq:fierz} we also obtain
\be
\tilde{\chi}_{1}\,\chi_{2}\,\epsilon=-\tilde{\chi}_{1}\,\epsilon\,\chi_{2}-\tilde{\chi}_{2}\,\epsilon\,\chi_{1}\ .
\ee

\newpage

%%%%%%%%%%%%%%%%%%%%%%%%%%%%%%%%%%%%%%%

\end{document}